\documentclass[fleqn,usenatbib]{mnras}
\usepackage{newtxtext,newtxmath}
\usepackage[T1]{fontenc}

\usepackage{graphicx}	
\usepackage{amsmath}	
\newcommand{\qap}{q_{\rm AP}}
\newcommand{\qrsd}{q_{\rm RSD}}
\newcommand{\mpch}{~h^{-1}{\rm Mpc}}
\newcommand{\msunh}{~h^{-1}{\rm M_\odot}}
\newcommand{\giturl}{\url{https://gitlab.com/dante.paz/popcorn_void_finder}}
\title[A cosmological test using the abundance of cosmic voids]{Guess the cheese flavour by the size of its holes: A cosmological test using the abundance of popcorn voids}

\author[D. Paz et al.]{
Dante J. Paz,$^{1,2}$\thanks{E-mail: dpaz@unc.edu.ar}
Carlos M. Correa,$^{1,3}$
Sebastián R. Gualpa,$^{1}$
Andres N. Ruiz,$^{1,2}$
Carlos S. Bederi\'an,$^{4}$
\newauthor
R. Dario Gra\~na$^{1}$
and Nelson D. Padilla$^{1}$\\
$^{1}$Instituto de Astronom\'ia Te\'orica y Experimental, UNC-CONICET, Laprida 854, X5000BGR C\'ordoba, Argentina\\
$^{2}$Observatorio Astron\'omico, UNC, Laprida 854, X5000BGR C\'ordoba, Argentina\\
$^{3}$Max-Planck-Institut f\"ur Extraterrestrische Physik, Postfach 1312, Giessenbachstr, D-85741 Garching, Germany\\
$^{4}$Instituto de F\'isica Enrique Gaviola, UNC-CONICET, Av. Medina Allende s/n, Ciudad Universitaria, X5000HUA C\'ordoba, Argentina
}

\date{Accepted XXX. Received YYY; in original form ZZZ}

\pubyear{2023}

\begin{document}
\label{firstpage}
\pagerange{\pageref{firstpage}--\pageref{lastpage}}
\maketitle

\begin{abstract}
We present a new definition of cosmic void and a publicly available code with the algorithm that
implements it. Underdense regions are defined as free-form objects, called popcorn voids, made 
from the union of spheres of maximum volume with a given joint integrated underdensity
contrast. The method is inspired by the excursion-set theory and consequently no rescaling
processing is needed, the removal of overlapping voids and objects with sizes below the shot
noise threshold is inherent in the algorithm.
The abundance of popcorn voids in the matter field can be fitted using the excursion-set
theory provided the relationship between the linear density contrast of the barrier and the threshold
used in void identification is modified relative to the spherical evolution model.
We also analysed the abundance of voids in biased tracer
samples in redshift space.  We show how the void abundance can be used to measure
the geometric distortions due to the assumed fiducial cosmology, in a test similar
to an Alcock-Paczy\'nski test.  Using
the formalism derived from previous works \citep{Correa2}, we show how to correct the abundance
of popcorn voids for redshift-space distortion effects. Using this treatment, in
combination with the excursion-set theory, we demonstrate the feasibility of void
abundance measurements as cosmological probes.  We obtain unbiased estimates
of the target parameters, albeit with large degeneracies in the parameter
space.  
Therefore, we conclude that the proposed test in combination with other cosmological probes has
potential to improve current cosmological parameter constraints.
\end{abstract}




\section{Introduction}

Of the variety of recognisable topological forms that make up the large-scale
structure of the Universe, cosmic voids are commonly regarded as vast
underdense regions surrounded by walls of filaments from which matter initially
escapes to later fall along the filaments and finally reach the galaxy
clusters.  Although this idea is conceptually behind most of the work in the
literature, making a precise definition of these regions in simulations and
observations relies on somewhat arbitrary criteria.  There are many void
finders in the literature and, equivalently, void definitions, each based on
some fundamental properties of voids.

A popular family of void finders are those based on feature analysis in the
matter density field \citep{platen_2007,ZOBOVpaper,VIDEpaper}.  The concept
here is to look for basins surrounded by bridges of matter.  This is the case,
for instance, of void finders based on watershed transformations, which by
making an analogy between the matter density field with topographic maps,
define voids as the regions where the falling water eventually drain.  They
have the advantage of assuming no shape for the void, as it is defined by the
basin surfaces naturally conforming to the filament walls surrounding the low
density regions.  This definition of a void, based on the density field
characteristics of the surrounding matter, is useful in some research contexts,
although it can also have some disadvantages in other contexts.  Some important
properties of the velocity field in voids are not guaranteed by this
definition, because the density field of matter within the void region is not
completely constrained.  In particular, the velocity field divergence within
these basins may be negative, rather than the expected positive behaviour, due
to the presence of high densities within these regions.  In other words, and
perhaps going too far in the analogy, the possible presence of islands within
basins can result in regions with a higher integrated density than is expected
for a void region.

Another criterion widely used in the literature is to define voids through its
dynamics, that is, using the characteristics of the tidal or velocity fields in
these regions \citep{hahn_2007,lavaux_2010,elyiv_2015}. Because voids are
underdense respecting to the average density of the Universe, they are expected
to expand at a faster rate than the rest of the Universe, which is often called
a super-Hubble expansion \citep{SvdW}.  Then voids regions can be thought as
zones of positive divergence in the velocity field. Equivalently, in
perturbation theory they can be associated with regions where the gravity tidal
field has only positive eigenvalues. Both type of criteria, based on velocity
or tidal field, can be used in simulations, defining physically meaningful
regions, however its application to galaxy surveys could be
somewhat difficult without the use of reconstruction methods to infer the tidal and
velocity fields from the galaxy positions in redshift space.

Finally, we will consider a third family of void finders, those that define a
void as an integrated underdensity in a given volume
\citep{hoyle_2002,colberg_2005,padilla_2005,brunino_2007,svfpaper}.  Generally
speaking, a parametric template shape is defined as a window function and then
convolved on the matter or tracer field, looking for regions with density below
to a given threshold. This is usually done by using a sphere of variable radius
centred somewhere (either randomly or at low-density local minima) and
calculating the total amount of matter or tracers within it. Voids are then
defined as the largest regions, generally non-overlapping each other, with a
negative density contrast below a certain parameter.  This class of finders are
usually referred to as spherical void finders.

As a consequence of Birkhoff's theorem, the negative density contrast within
the spherical region ensures that the outer spherical shell will expand at a
rate greater than the universal average \citep{1993Peebles}. Therefore it can
be thought that this void finder methodology is somewhat closer than others to
theory of void abundances developed in \cite{SvdW}. The convolution in space of
spheres of different radii resembles the excursion-set framework while the
constraint on integrated density relates closely to the barriers in this theory
(of course leaving aside the differences between the nonlinear fluctuations
field and the initial Gaussian field).  However, as we shall see, due to the
imposed spherical shape of the window function some underdense regions can be
artificially identified as multiple spherical voids as consequence of its
intrinsic shape.

Beyond the three types of void definitions presented in this introduction
(which is not intended to be exhaustive), there are many more algorithms
available to define and identify voids in simulations and observations.  Given
the zoo of available methods to define a void, it is worth asking which are the
best criteria to follow when choosing one.
A pragmatic choice is to use the definition that allows us to extract the
greatest meaning from void samples in a given research context. 
In this work, we will focus on the abundance of cosmic voids as a cosmological probe.
In other contexts, the
interested reader can check the different comparison projects as
\citet{colberg_comparison_2008}, \citet{cautun_comparison_2018} and
\citet{paillas_comparison_2019}. 

Several methods using voids have been developed to measure background model
parameters, test dark energy models, and constrain alternative theories of
gravity.  These methods can be classified into two main statistics: the
void-galaxy correlation function (hereafter VGCF) and the void size function
(hereafter VSF).

On small to intermediate scales, the VGCF can be interpreted as the average
density profile of galaxies, or more generally matter tracers, around void
regions.  This is considered in what we call \textit{1-void term} scales,
following \citet{Cai16}, in analogy with the halo model of the correlation
function.  The behaviour at larger scales, that is, in the \textit{2-void
term}, is given by the clustering of the tracer field multiplied by a bias
factor that describes the clustering of voids at these scales \citep{Cai16}. By
correctly modelling this function and its measurement in galaxy surveys, see for
instance \citet{clues2,Cai16}; \citet{rsd_achitouv2,rsd_chuang,rsd_hawken1,rsd_hawken2};
\citet{rsd_achitouv3,rsd_nadathur,Nadathur_erratum,rsd_hamaus,Hamaus20,
APeuclid_hamaus,2022Woodfinden} and references in this last work,
it is possible to perform an Alcock-Paczy\'nski test
\citep[hereafter AP test, ][]{AP} similar to the cosmological tests based on
the two point correlation function at scales of the Baryon Acoustic Oscillation
(BAO) \citep[for BAO tests see for instance][]{Sanchez17}.
In particular, in a previous work \citep{Correa1}, we developed a new test
based on two perpendicular projections of the VGCF with respect to the
line-of-sight direction.  Our method provides three novel aspects.  First, we
propose a fiducial-free test, since correlations are treated directly in terms
of void-centric angles and redshifts, without the need of explicitly assuming a
distance scale.  Second, the combination of working in this observable-space
framework with two perpendicular projections of the VGCF allows us to
effectively break any possible degeneracy in the parameter space.  Finally, the
covariance matrices associated with the method allow us a robust Bayesian
analysis and to significantly reduce the number of mock catalogues needed to
estimate them.

However, some difficulties arise in modelling cross-correlations between voids
and galaxies in redshift surveys \citep{Correa3}. The redshift-space 
distortions due to the tracer velocity field induce apparent patterns in the
distribution of galaxies around voids depending on their intrinsic dynamics.
Some void regions are surrounded by large overdensities, resulting in void
overcompensation and, as a consequence, large-scale region shrinkage. On the
other hand, some voids are surrounded by underdense or slightly overdense
regions, achieving large-scale asymptotic compensation, so these regions tend
to expand at all scales. Thus, depending on the large-scale void environment,
we have different modes of void evolution, often called \textit{void-in-void}
and \textit{void-in-cloud}, as described in \cite{SvdW} and first observed in
redshift surveys in \citet{clues2}. This phenomelogy complicates the control of
systematics in VGCF cosmological tests.

In this work we will focus on the analysis of the void size function 
\citep[][]{abundance_furlanetto,abundance_achitouv,
abundance_pisani,2016Pollina,bias_pollina1,bias_pollina2, abundance_ronconi2,abundance_bias_contarini,
ContariniArxiv1,ContariniArxiv2,Euclid_forecast,abundance_verza,2022VerzaA,2022VerzaB,Correa2}
and its possible exploitation in the determination of cosmological parameters
in galaxy surveys. 
This function describes the abundance of voids of a given
size and can be modelled using the excursion-set formalism combined with the
spherical evolution of matter underdensities derived from perturbation theory
\citep{SvdW,2013Jennings}. 

The \citet{SvdW} formalism was implemented in void
statistics by assuming that voids form in isolation due to only their initial
density. \citet{2013Jennings} have shown that, as a consequence of this 
assumption, the model seems to overestimate the abundance of voids. 
These authors obtain a significantly smaller number of voids by
imposing the condition that voids occupy a constant fraction of the volume of
the Universe. Without this assumption, the \citeauthor{SvdW} model results
in a fraction of the volume of the Universe occupied by voids greater than unity,
after the transition to nonlinear regime.
See Section~\ref{sec:vsf_model} for more details regarding the VSF modelling.
\citet{2013Jennings} also compare models with samples of voids identified in the matter field
using the \textsc{zobov} void finder \citep{ZOBOVpaper}, which employs a watershed
transformation approach to identify void regions. 
However, to obtain abundances of voids in simulations comparable to those
predicted by the models, it is necessary to ensure that each voids encloses a
region with an integrated density contrast below the desired
threshold. 
Voids that do not satisfy this condition should be discarded.
Given this approach, \citet{2017RonconiMarulli} provides the
implementation of a set of numerical tools for analysing cosmic void catalogues,
implemented within \textsc{CosmoBolognaLib} \citep{2016Marulli}. This toolkit
facilitates the implementation of cosmological tests based on void abundances.
These authors also confirm that in the case of \textsc{vide} voids 
\citep[][this finder is a modification of the \textsc{zobov} code]{VIDEpaper}, the
cleaning procedure to make theory and data comparable requires removing almost
$90\%$ of the objects in the raw sample of voids. After this step, it is
necessary to scale the void sizes to meet the density threshold
required by theory. By definition, these procedures have no effect when
applied to voids identified in the integrated density contrast 
\citep[see for instance][]{svfpaper,Correa2}. We verified in the case of our
spherical void finder the procedures of \citet{2017RonconiMarulli} barely 
remove or resized objects, leaving the catalogue almost identical.

In a previous work \citep{Correa2}, we show how the abundance distribution of
voids identified in the integrated density contrast can be used to perform an 
AP test. We derive analytical correction factors on void sizes for geometric
distortions (GD) due to the assumption of a fiducial cosmology and for redshift
space distortions (RSD) due to tracer dynamics around voids. These formulas
allow us not only to correct for the effect of redshift-space distortions, but
also to add an additional dependency on cosmological parameters with respect to
excursion-set theory that can be exploited in the tests. Briefly, the void 
sizes can be corrected for distortions using linear theory and thus this adds a
dependence on the growth rate of the structures. On the other hand, the
geometric distortions add a dependency on the parameters of the background
model and therefore with the expansion history of the Universe.

In this work we derive a new definition of a void: the popcorn void finder.  As
we will see in Section~\ref{sec:abundances}, this void definition has some
advantages that make it useful in the context of void abundance studies and
comparison with models.  The algorithm, presented in
Section~\ref{sec:algorithm}, has an important and distinctive feature: voids
are defined as free-form objects by means of the integrated density contrast of
underdense locations of space.  We also release open source software to
identify popcorn voids in cosmological simulations\footnote{\giturl}. In
Section~\ref{sec:abundances}, using the correction factors for the redshift and
geometric spatial distortions derived from \citet{Correa2}, and excursion-set
models for the abundance of voids, we implement a cosmological test for the
abundance of popcorn voids identified in the redshift space over biased tracer
samples. Finally, in Section~\ref{sec:discus} we discuss our results in the
current context of cosmic void abundance studies.

\section{Void size function modelling}
\label{sec:vsf_model}

The void size function (denoted here with $A_\mathrm{v}$),
depict the void abundance as the number density of this objects
in a given logarithmic interval in void sizes divided by the 
length of this interval, that is:
\begin{equation}
    A_\mathrm{v}(R_{\rm v}):=\frac{d\mathrm{n_v}}{d\mathrm{ln}R_{\rm
	v}}=\frac{1}{V}\frac{d\mathrm{N_v}}{d\mathrm{ln}R_{\rm v}}\,,
    \label{eq:VSF}
\end{equation}
where $d\mathrm{N_v}$ is the total number of voids with the natural logarithm
of their radius, $\mathrm{ln}(R_{\rm v})$, between $\mathrm{ln}(R_{\rm v})$ and
$\mathrm{ln}(R_{\rm v})+d\mathrm{ln}R_{\rm v}$, while $V$ is the volume
of the universe where voids are studied. 

The excursion-set formalism is generally used to model the void size
function as the halo mass function as well. Briefly, this formalism is based
on the fact that over the initial Gaussian field, the integrated density contrast
$\Delta$ on a sphere of radius $s_m$ describes a Markovian process as $s_m$ increases.
The jump probability in such a process depends on the power spectrum of the density field,
therefore also on the cosmological parameters. Generally speaking, if such a
sphere centred at a given random position in the initial conditions contains
more matter than a given critical density contrast, the matter inside it is
expected to collapse and form a virialised object \citep{PressSchechter}. 
Similarly, if the sphere contains a negative density contrast, below a given
threshold (void barrier), and do not cross at larger scales the collapse
barrier, it is likely to develop a void region. 

In this way, it is possible to relate the classical hypothesis tests for
crossing barriers in Markovian processes with the statistics of rare peaks 
(galaxy clusters) or density minima (voids). In other words, by counting the
number of processes that cross certain barriers in the initial conditions, 
it is possible to infer the abundance of structures at present time. 

With this approach, and considering the definition of Eq.~(\ref{eq:VSF}),
\cite{SvdW} derived the following model for the VSF:
\begin{equation}
    A_{\rm v}(R_{\rm v}) =
    \frac{f_{\mathrm{ln} \sigma}(\sigma)}{4/3 \pi (R_{\rm v}^{\rm L})^3} 
    \frac{d \mathrm{ln} \sigma^{-1}}{d \mathrm{ln} R_\mathrm{v}^{\rm L}},
    \label{eq:vsf_svdw}
\end{equation}
referred to as the SvdW model.  Here, $\sigma$ is the square root of the mass
variance, defined in terms of a smoothing scale $\mathcal{R}$ as follows:
\begin{equation}
    \sigma^2(\mathcal{R}) = \int \frac{k^2}{2 \pi^2} P_m(k) |W(k, \mathcal{R})|^2 ~ dk,
\end{equation}
where $P_m(k)$ is the matter power spectrum, $W(k, \mathcal{R})$ a filter
function, and $f_{\mathrm{ln}\sigma}$ the fraction of the Universe occupied by
voids:
\begin{equation}
    f_{\mathrm{ln}\sigma} = 2 \sum_{j=1}^{\infty} 
	j\pi \chi^2 \mathrm{sin}(j\pi \mathcal{D}) \mathrm{exp} 
	\left[ - \frac{(j\pi \chi)^2}{2} \right],
    \label{eq:flnsig}
\end{equation}
where in turn, $\chi = \mathcal{D}\sigma / |\Delta_{\rm v}^{\rm L}|$ and
$\mathcal{D} = |\Delta_{\rm v}^{\rm L}| / (|\Delta_{\rm v}^{\rm L}| +
\Delta_{\rm c}^{\rm L})$.  $\Delta_{\rm c}^{\rm L}$ and $\Delta_{\rm v}^{\rm
L}$ represent the two barriers needed in the excursion-set to take into account
both the void-in-cloud and void-in-void modes.  The former is expected to vary
within $1.06 < \Delta_{\rm c}^{\rm L} < 1.686$, the moments of turn-around and
collapse, respectively.  The latter is expected to be associated with the
moment of shell crossing in the expansion process.  
Therefore a key ingredient is the relationship between the linear
barrier $\Delta_{\rm v}^{\rm L}$ and the non-linear contrast used to identify voids
in numerical simulations, namely $\Delta_{\rm v}$.
This relationship is typically established based on the spherical evolution
model, which initially assumes small density perturbations that evolve
isotropically towards a non-linear regime embedded in a FLRW background.
Finally, $R_{\rm v}^{\rm L}$ is the linear radius predicted by theory.  It can be related to its
non-linear counterpart by a constant factor: $R_{\rm v} = \gamma R_{\rm v}^{\rm
L}$, with $\gamma = (1 + \Delta_{\rm v})^{-1/3}$, which follows from
considering the radius of the underdense region in the spherical expansion model at the moment of shell
crossing, i.e. when it reaches the average non-linear density contrast of
$\Delta_{\rm v}$.  In this sense, $\Delta_{\rm v}^{\rm L}$ is the
linearly-extrapolated value of $\Delta_{\rm v}$.

The key assumption of the SvdW model is that the comoving number density of
voids is conserved during the evolution.  However, this assumption leads to a
cumulative comoving volume fraction in voids that exceeds unity.  To fix this
problem, \cite{2013Jennings} suggest that the comoving volume fraction must be
fixed during the evolution, instead.  In this picture, when a void evolves, it
combines with its neighbours to conserve volume and not number.  In this way,
the abundance of voids becomes
\begin{equation}
    A_{\rm v}(R_{\rm v}) = 
    \frac{f_{\mathrm{ln}\sigma}(\sigma)}{4/3 \pi R_{\rm v}^3} 
	\frac{d\mathrm{ln}\sigma^{-1}}{d\mathrm{ln}R_{\rm v}^{\rm L}}
    \frac{d \mathrm{ln} R_{\rm v}^{\rm L}}{d \mathrm{ln} R_{\rm v}},
    \label{eq:vsf_vdn}
\end{equation}
referred to as the volume-conserving (Vdn) model.
Note that a simple comparison between Eqs.~(\ref{eq:vsf_svdw}) and
(\ref{eq:vsf_vdn}) indicates that $A_{\rm v} [{\rm Vdn}] = A_{\rm v} [{\rm
SvdW}]$ / $\gamma^3$.  Therefore, the SvdW and Vdn models only differ by a
constant amplitude.  

In practice, we can only define voids from the observed galaxy
distribution.  In this case, it is not trivial to relate linear density barrier to a
specific density threshold. This can be overcome by the procedure suggested by
\cite{abundance_bias_contarini}.  First, we define an observational density
threshold as low as possible: $\Delta_{\rm v, g}$ (usually less than $-0.7$, in
this work we use $-0.9$).  Then, we relate this value to the one corresponding
to the total matter density field by means of the bias parameter: $\Delta_{\rm
v} = \Delta_{\rm v, g}/b_\mathrm{eff}$.  Finally, we derive its corresponding
linear value with the fitting formula provided by \cite{1994Bernardeau}:
$\Delta_{\rm v}^{\rm L} = C[1-(1+\Delta_{\rm v})^{-1/C}]$, with $C=1.594$.

A last, but very important aspect, must be considered in the VSF modelling.
The models presented up to now make predictions for void radii $R_{\rm v}$ in
real space.  Following \cite{Correa2}, there is a linear relation between the
observational (redshift space) void radii $\Tilde{R}_{\rm v}$ and its
real-space counterpart by means of the AP and RSD factors given by
Eqs.~(\ref{eq:qap}) and (\ref{eq:qrsd}), respectively:
\begin{equation}
    \Tilde{R}_{\rm v} = q_{\rm AP} ~ q_{\rm RSD} ~ R_{\rm v}.
    \label{vsf_zspace}
\end{equation}
As we shall see in the folloing sections, the factors introduced here encode
the AP-volume and RSD-expansion effects that voids suffer when they are mapped
from real to redshift space, and therefore, encode valuable cosmological and 
dynamical information.  We highlight the importance of considering this step
in order to obtain unbiased cosmological constraints when designing cosmological tests.

\section{The popcorn void finder}
\label{sec:algorithm}

\begin{figure}
	\includegraphics[width=\columnwidth]{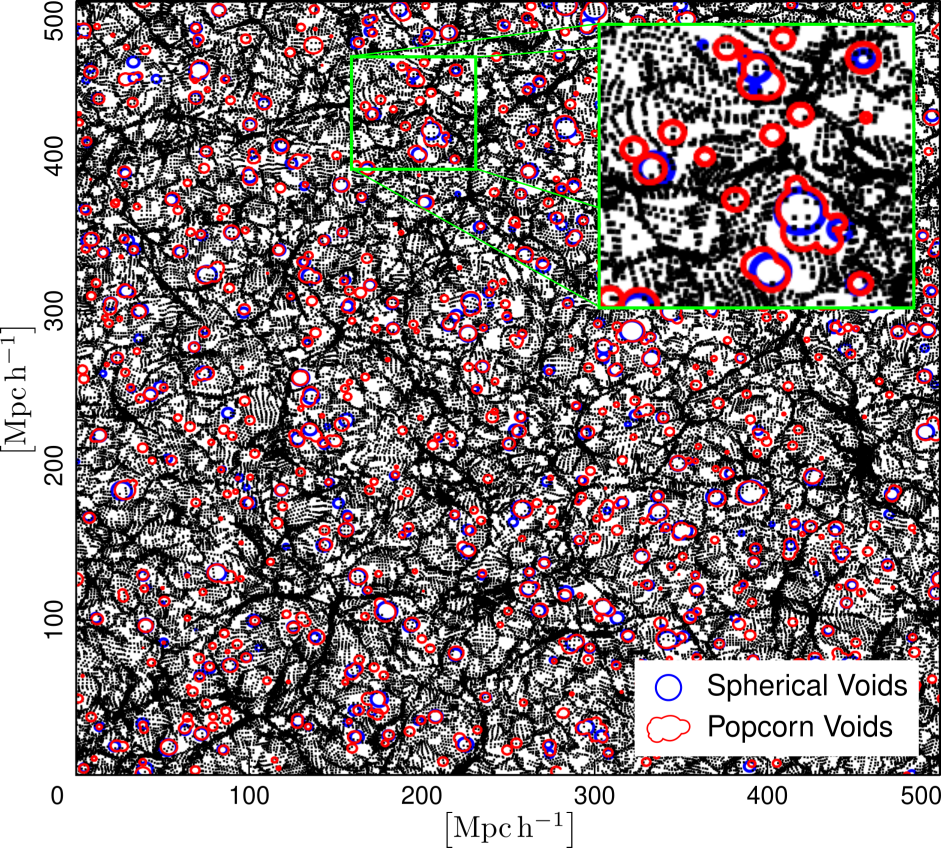}
    \caption{
    Slice of a dark matter-only simulation at final time (sim-1, $512^3$ particles in a box of $500\mpch$ side length). The particles (black dots) in the x-y plane have a z-coordinate between $245-255\mpch$.
    Integrated density voids (with $\Delta \leq -0.9$) were identified on the box assuming spherical (blue circles) and free-form popcorn shapes (solid red lines, see text for detailed description). These objects are shown by their intersection with the x-y plane at z$=250\mpch$ (solid line in red or blue).
    }
    \label{fig:slice}
\end{figure}

As we briefly described in the introduction section, there are several void
finders in the literature, each based on some physical characteristic of these
regions. In this section we introduce a new void finder, designed to improve
void abundance analyses to some extent. Roughly speaking, the basic idea behind this
approach is to look for two desired features of a void object\footnote{ We use
the term void object to name the output of a void finder that can eventually
be associated with an underdense region, that is, the physical region of
interest that we seek to identify and characterise.}. 

The first feature we look for is to obtain a definition with some kind of free
form, that is, the shape of the void object must be flexible enough to fill the
entire underdense region. As we will see in the following subsections, the
intention here is to avoid void fragmentation, that is, to identify multiple
void objects in association with a single physically significant underdense
region in the large-scale distribution. This also allows us to associate a single
scale to the void region, using for instance the equivalent radius of a sphere
with the same volume than the void object.

The other important aspect is to identify regions that enclose a well-defined
integrated density contrast. Therefore at the moment of identification the
density contrast can be taken to be below some desired threshold. This
parameter could be then related with a density barrier in excursion-set models
at a given scale. In the following subsections, we describe in detail the
popcorn void finder algorithm (Section~\ref{subsect:algorithm}) and its
application in the dark matter distribution of $\Lambda$CDM cosmological
simulations (Section~\ref{subsect:prop}). We restrict ourselves to the
analysis of simulations, in this first version of the algorithm, because the
periodic boundary conditions of this type of data allow us to avoid the
treatment of more complicated selection functions, such as those intrinsic to
galaxy surveys. In this way a popcorn object is allowed to fill an underdense
region without a boundary limitation.  We are leaving for future development a
version of the popcorn void finder that can be run on galaxy surveys,
implementing similar methods to those used in the spherical void finder on
observational data \citep[see for instance][]{2022Alfaro,2022Rodriguez}.

Regarding the simulations in this
work we analyse three boxes: sim-1, sim-2 and sim-3, with particles of $512^3$,
$1024^3$ and $6720^3$ respectively in periodic boxes of $500$, $1000$ and 
$3000$ $\mpch$ of side length. All boxes are periodic realisations
of a flat $\Lambda$CDM cosmology with the matter density and dark energy
parameters given by $\Omega_m=0.25$, $\Omega_\Lambda=0.75$, respectively and
a root mean square variance of linear perturbations of $\sigma_8=0.9$.
The sim-3 box is in fact the Millennium XXL simulation
\citep[MXXL, to see details about this run please read][]{AnguloMXXL}.

\subsection{Free form and integrated underdensity void definition}
\label{subsect:algorithm}
\begin{figure}
	\includegraphics[width=\columnwidth]{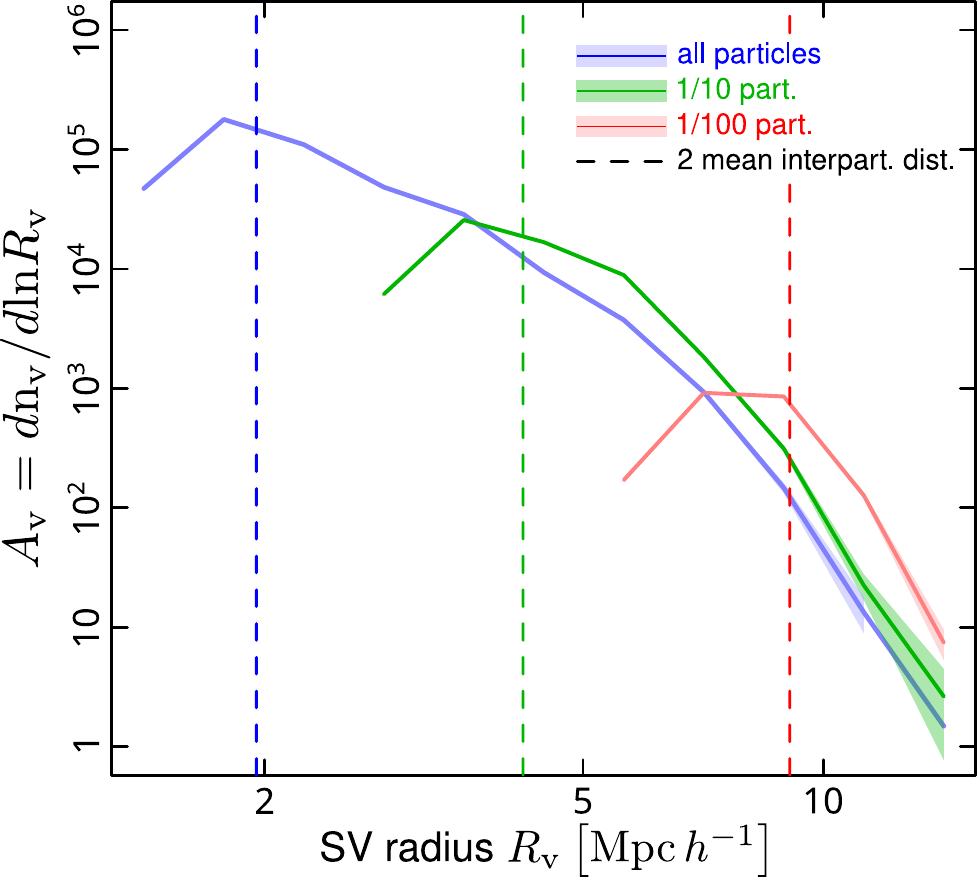}
    \caption{
    Abundance of spherical voids identified in the dark matter field of the sim-1 box 
    (see Sect. \ref{sec:algorithm}) as a function of their radius (blue). 
    In green and red we show the abundances when
    only 1/10 and 1/100 fractions of randomly selected particles are used. 
    Shaded areas correspond to shot-noise error bars.
    The dashed vertical lines correspond to 2 times the mean distance between
    particles for these three different number densities.
    }
    \label{fig:testshot}
\end{figure}

In order to compute the integrated density of a given region, it is necessary
to define a three dimensional boundary that defines the domain to perform
such integration. We define the integrated density of a given region as the
total amount of matter or tracers inside divided by its volume. In this work we
distinguish this definition with respect to the most used of the differential
density field, or simply the density field $\rho$, defined by 
$d\mathrm{M}=\rho(\mathbf {x})dV$, where $d\mathrm{M}$ is the total amount of
matter in a differential volume $dV$ at a given position $\mathbf{x}$ in space.
This latter scalar field can be estimated on the data using various methods,
from a regular fine mesh of the simulation volume, a tessellation of the tracer
distribution, or particle kernel interpolation, among others.

The popcorn void finder approach, as we will see, is a generalisation of the
spherical void finder (hereafter SVF) by adding in a recursive way more spheres
to fill the void region. This approach is similar to those used in the works of
\citet{colberg_2005} and more recently \citet{Douglass_2022}, although with
some differences. We start by identifying spherical voids (SV) on integrated
density in the simulation box, following a procedure similar to that described
in \citet{svfpaper}. Briefly, after seeding the simulation box in regions of
local low density, we place spheres in each seed and increase their radius as
much as possible to keep an integrated density contrast below a certain
threshold, $\Delta_\mathrm{v}$. After this, all overlapping spheres are removed
keeping the largest ones. In this paper and in the accompanying published code,
we follow a slightly different algorithm than our previous articles; however,
it follows the same two main steps mentioned above. As a consequence, the
spherical void samples in this new version do not change significantly from our
previous results. Their abundance and distribution in space are very similar,
however the new SVF provided in this work performs significantly faster and
allows us to identify spherical voids in simulations with higher resolutions.
For a detailed description of the method, see Appendix~\ref{svf_algorithm}. We
also provide the SVF algorithm used here within the public code released along
with this work.

In Fig.~\ref{fig:slice} we show a slice of the spatial distribution of dark
matter particles in the sim-1 box at $z=0$. The slice depth is $5\mpch$, while
all spherical voids found on the particle distribution and intersecting the
mid-plane of the cut are shown using their intersecting circle in blue solid
lines. These SV objects are defined with an integrated density contrast below
the threshold of $\Delta_\mathrm{v}=-0.9$. On the same figure, we also over
plot popcorn voids in red, as we will introduce later in this subsection. As
stated before, SVs are the first step in defining popcorn objects, so it is
expected that there will be at least one of these objects at every SV location. 

\begin{figure*}
    \centering
    \includegraphics[width=2\columnwidth]{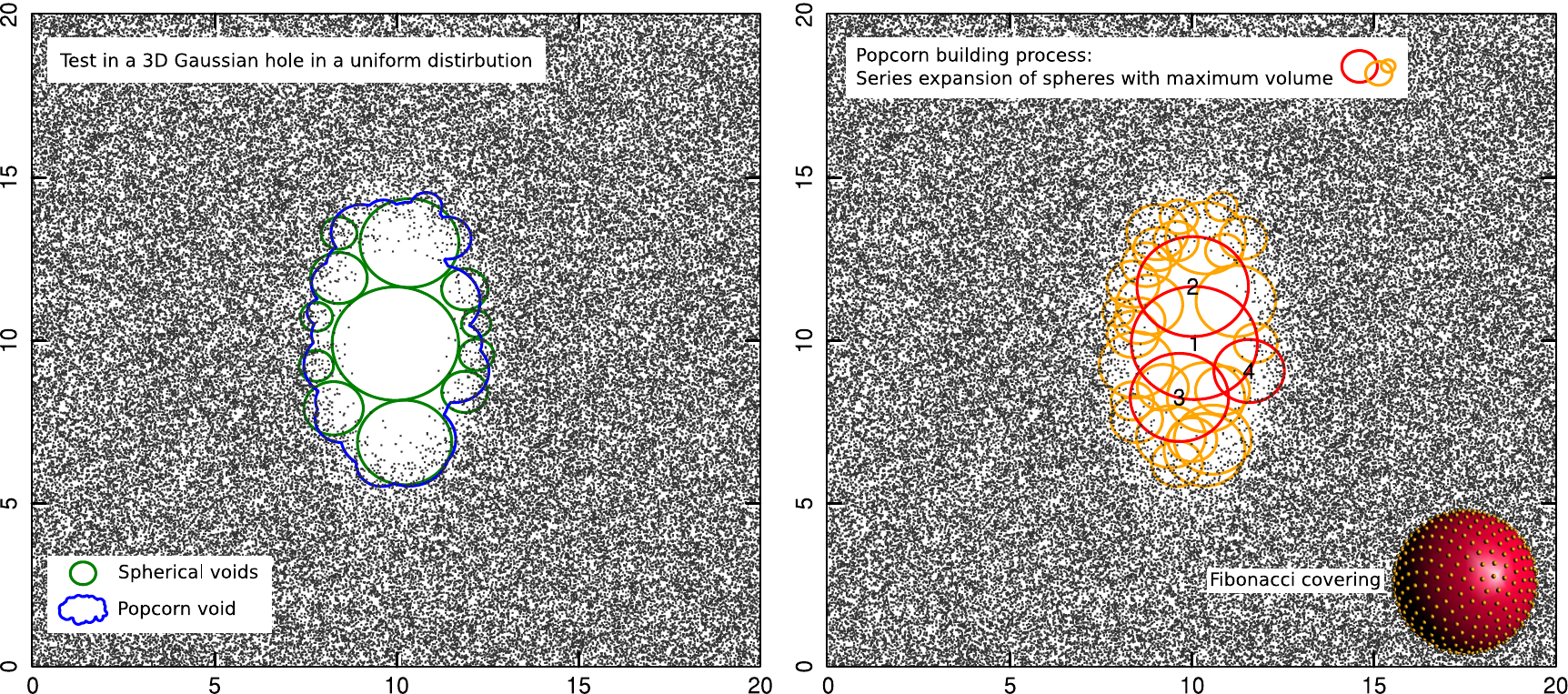}
    \caption{X-Y slice at Z$=10$ of a ellipsoidal Gaussian hole in a uniform
	random distribution of points in a cube of side 20 (dimensionless).
	Some points were drawn from the uniform box with a probability given by
	the product of three normal distributions over their Cartesian
	coordinates (see details in the text).
	\textit{Left panel:} Spherical and popcorn void objects shown by their
	intersection with the midplane at Z$=10$ (green circles and solid blue
	curve, respectively, as indicated by the key). \textit{Right panel:}
	The member spheres of the popcorn object (orange circles). The first 4
	spheres added by the algorithm are indicated by red circles and
	labelled by their inclusion order.  The bottom right inset is an
	example of a Fibonacci covering of a sphere (see the text)} 
    \label{fig:cartoon}
\end{figure*}

The SV abundance in the sim-1 simulation as a function of sphere
radius is presented in Fig.~\ref{fig:testshot}, in solid blue line, 
as defined in Eq. \ref{eq:VSF}. 
We also present in green (red) the abundances obtained when
only a particle of ten (one hundred) is used randomly chosen in the SV
identification. As can be seen, void sizes tend to be larger in these random
diluted samples compared to the full sample. Random particle dilution more
efficiently suppresses shortwave fluctuations in the matter field, erasing
small scale structures. In this way, small voids are incompletely detected and
merged into larger regions.  The corresponding transparent shaded areas are the
error bands. We also indicate with vertical dashed lines, the scale of twice
the mean interparticle distance. The behaviour of these abundances is
qualitatively consistent with what we expect in a hierarchical clustering
scenario: small voids outnumber larger ones, a trend similar to a Schechter
function predicted by excursion-set theory. However, at radii less than a few
times the mean interparticle distance, the abundance shows the opposite
behaviour. This is expected due to the shot noise dominance at low density and
low scale regions. This figure helps us to define what we call the shot noise
radius, $R_\mathrm{shot}$, which is the minimum radius for which spherical
voids have physical meaning  (see, for instance, the analysis presented in
\citet{Correa2}).  Therefore, this radius will be crucial in defining the
popcorn voids.

As anticipated, the next step in the popcorn algorithm is to add more spheres
to closely cover the void region. This is achieved by placing seeds on the
surface of the spherical object. The seed locations follow the Fibonacci
covering lattice. In an inset on the right panel of Fig.~\ref{fig:cartoon} we
show an example of seed distribution (see the small orange dots on a red
sphere). As can be seen in the figure, the seeds are placed regularly aligned
on a double spiral structure on the surface of the sphere (the so called Golden
Spirals). This covering scheme has some advantages over regular lattices, also
being found very frequently in nature \citep[e.g. sunflower seeds,  pineapple
scales, spikes in the \textit{cactus Mammillaria}, see][]{Gonzalez2009}.  On
the one hand, the total number of points in the scheme can be set to any odd
natural number, unlike regular distributions that increase with the square
\citep[see for example][]{healpix,Gonzalez2009}, this allows us to smoothly
vary the number of seeds with the radius of the sphere to be covered, to ensure
the desired spatial resolution (larger spheres will be more covered by
seeds)\footnote{However, we have found that there are little to no differences in the 
shape and scale distribution of void objects when using a fixed number of
seeds while keeping at least one seed per $0.1$ steradian. This setting is 
preferable to minimise computing time in large voids.}
The other important advantage is that the seeds in the Fibonacci lattice are
optimally packed \citep{ridley1982,ridley1986}, sampling the sphere regularly
while each point has a different latitude, which allows a better sampling than
regular lattices \citep{Gonzalez2009}. This last characteristic is crucial to
obtain a correct coverage of the void regions, regardless of their orientation
and shape.

After covering, each seed expands one at a time, seeking its maximum radius to
satisfy that the integrated density in the volume of the union of the two
spheres is below the threshold, $\Delta_\mathrm{v}$. Then after considering
each seed, the one that has expanded the most is accepted. The next step is to
cover the joining surface of these two spheres, that is the surface of the
initial spherical void and the accepted sphere. This is done by covering each
sphere with a number of seeds appropriate to its size and the desired spatial
resolution, and not considering those seeds that fall inside any sphere.  Then
the process of seeding, keeping only the most expanded sphere, and seeding
again is repeated iteratively. Thus, in the second step, the most expansive
seed in the joint volume of three spheres is accepted, and in subsequent steps
the volume of the union of four, five spheres, is taken into account. The
process ends once no sphere with a radius greater than a given threshold
$R_\mathrm{th}$ can be added, satisfying the integrated density contrast
condition ($\Delta < \Delta_\mathrm{v}$). This radius threshold is chosen as
the radius of the shot noise, $R_\mathrm{th}=R_\mathrm{shot}$.

In this way it can be deduced that the measurement of $R_\mathrm{shot}$ is
fundamental for the definition of a popcorn object. $R_\mathrm{th}$ can be
defined a priori as a few times the mean interparticle distance or it can be
chosen more precisely as the radius where the behaviour SV abundance change.
That is, $R_\mathrm{shot}$ can be defined as the scale at which the SV
abundance departs from a power law with negative slope. We have found that if
spheres with radius smaller than $R_\mathrm{shot}$ are allowed, artificially
large void objects are obtained, with a complex multiconnected topology. As a
side effect, the total computation time becomes significantly longer. On the
other hand, a large value of this threshold ($R_\mathrm{th}> R_\mathrm{shot}$)
results in artificially fragmented voids, that is multiple objects associated
to a single underdense region, and at small scales popcorn objects are
identical to their associated spherical void. 

In Fig.~\ref{fig:abundances_matter} we show the abundances of spherical voids
(blue) and popcorn voids for two different values of $R_\mathrm{th}$ (red and
green) identified over the sim-2 particle distribution at $z=0$. In an analogous
way to the case of spherical voids, the VSF defined in Eq.~(\ref{eq:VSF}) is
generalised to popcorn voids by replacing the sphere radius by an equivalent
radius defined as the radius of a sphere $R_\mathrm{v}$ of equivalent volume to
that of the popcorn void ($V_\mathrm{v}$),that is:
\begin{equation}
    R_\mathrm{v}=\left(\frac{3\,V_\mathrm{v}}{4\pi}\right)^{1/3}
    \label{eq:eqvrad}
\end{equation}
The shaded areas in Fig.~\ref{fig:abundances_matter} correspond to one standard
deviation around the measurements. The solid green line is the abundance of
popcorn objects obtained using a larger than ideal threshold radius, that is
$R_\mathrm{shot} < R_\mathrm{th}=5\mpch$.  In this case, the abundance of
popcorn objects closely matches that of spherical ones up to $R_\mathrm{th}$,
as expected by definition. After this scale, the abundance shows a behaviour
different from that expected from the abundance models, having an inflection
point close to the chosen threshold radius. The solid red line corresponds to
the abundance of popcorn voids with $R_\mathrm{th}=R_\mathrm{shot}$. 
The abundance of these objects behaves qualitative similar to what is
expected for the theory. However as can be seen in the figure, their abundance
is by far underpredicted by the  excursion-set models of \citet{2013Jennings} (Vdn), 
when a linear density barrier of $\Delta_{\rm v}^{\rm L}=-5.16$ is used, that is, the corresponding value
on the spherical evolution model for a non-linear density threshold
of $\Delta_\mathrm{v}=-0.9$ (similar behaviour is obtained using $\Delta_\mathrm{v}=-0.8$).
In the next section we will return to these results and their comparison with
theoretical predictions of void abundances. Finally, after running the
iterative process described above for each SV, all overlapping popcorn
candidates, with an intersecting volume larger than the volume
of the shot noise sphere ($4\pi/3 R_\mathrm{shot}^3$), are removed 
starting with those with the smallest initial SV and
keeping those with the larger initial SV.

The popcorn void finder algorithm can be summarised in the following recipe:

a) Seed regions of local low density and grow spheres to the largest radius
allowed by $\Delta < \Delta_\mathrm{v}$.

b) Remove all overlapping spheres starting with the smallest ones and keeping
the largest ones. In this step, a catalogue of spherical voids is produced,
each of which is taken as a popcorn candidate.

c) Cover regularly the surface of the popcorn candidate with seeds.

d) Expand each seed to the largest radius that the condition 
$\Delta < \Delta_\mathrm{v}$ allows in the joint volume of the candidate popcorn
and the tested sphere.

e) The seed that expands the most and has a radius greater than
$R_\mathrm{shot}$ is accepted and the current popcorn candidate is updated.

f) The process is repeated iteratively from step c) until no sphere can be
added to the candidate.

g) All intersecting popcorn objects are removed, {\rm tolerating volume
intersections lower than the shot noise limit, starting with the candidate with
the smallest initial SV and keeping those with larger initial SV.}

In Fig.~\ref{fig:cartoon} we present an illustrative example of the application
of the method described above. In a three-dimensional uniform random
distribution of points, we have eliminated some of them according to a 3D
multivariate normal distribution. The three joint Gaussian probabilities of
this distribution have the same mean, that is, the coordinates of the centre of
the box and different second moments. In this way we have obtained a kind of
triaxial Gaussian hole, as can be seen in the slice shown in the figure. This
toy data set allowed us to test some aspects of the popcorn algorithm as we
designed it. In the right panel of Fig.~\ref{fig:cartoon} we show a section of
the spherical members of the resulting popcorn after the step g (solid orange
and red lines). The red circles correspond to a slice of the spheres added in
the first four iterations of the algorithm, in the order labelled by the
numbers in the figure. In blue on the left we show the corresponding
intersection of the popcorn surface and with green circles the identified
spherical voids. This is a simple example of void region fragmentation when
using a spherical void finder. As we will see in the following sections, this
is an important aspect in exploiting void abundances as cosmological tests. In
this data set, the regions of maximum volume below any density threshold are,
by definition, ellipsoids at the centre of the box with semi-axes proportional
to the second moment of the Gaussian distributions. The integrated density
contrast in these ellipsoids can be derived using the corresponding error
functions of the three axes. We have also tested the effect of the angular
orientation of the multivariate distribution on the void finder results.
Regarding volume estimation, we have found that the popcorn void finder
converges uniformly to the expected values depending on the minimum allowed
sphere size and also on the numerical density of random points. This
convergence is faster using the Fibonacci covering scheme than the regular seed
distribution for all orientations.

To improve the clarity of the above description of the popcorn void finder, a
non-trivial technicality has been delayed until now. To the authors' knowledge,
there are no simple analytical formulas for the joint volume of more than three
spheres. The popcorn algorithm requires thousands of volume calculations of
arbitrary distributions of spheres in the assembly of each popcorn candidate.
Calculation of volumes by Monte Carlo methods is not feasible here due to
computation time and numerical precision required. The problem of calculating
the joint volume of a union of spheres is a well-known problem in the field of
Chemical Physics. For instance, the estimation of the solvation energy of
proteins usually requires calculation of the surface area and volume of
macromolecules, which are modeled as the superposition of a large number of
spheres (atoms). The \textsc{arvo} \citep{ARVO,ARVO2} package solves this problem
through an extremely efficient analytical approach, calculating the surface
area and volume of arbitrary sets of overlapping spheres with high precision.
We use the C version of the library provided by the authors of this software.

In this subsection, we have provided a full description of the algorithm used
in the definition of popcorn voids; for more details, the interested reader can
review the documentation and source code provided on \giturl. In the next
subsection we provide some examples of the application of this new void finder
in numerical simulations.

\subsection{Popcorn void properties on matter field}
\label{subsect:prop}

\begin{figure}
	\includegraphics[width=\columnwidth]{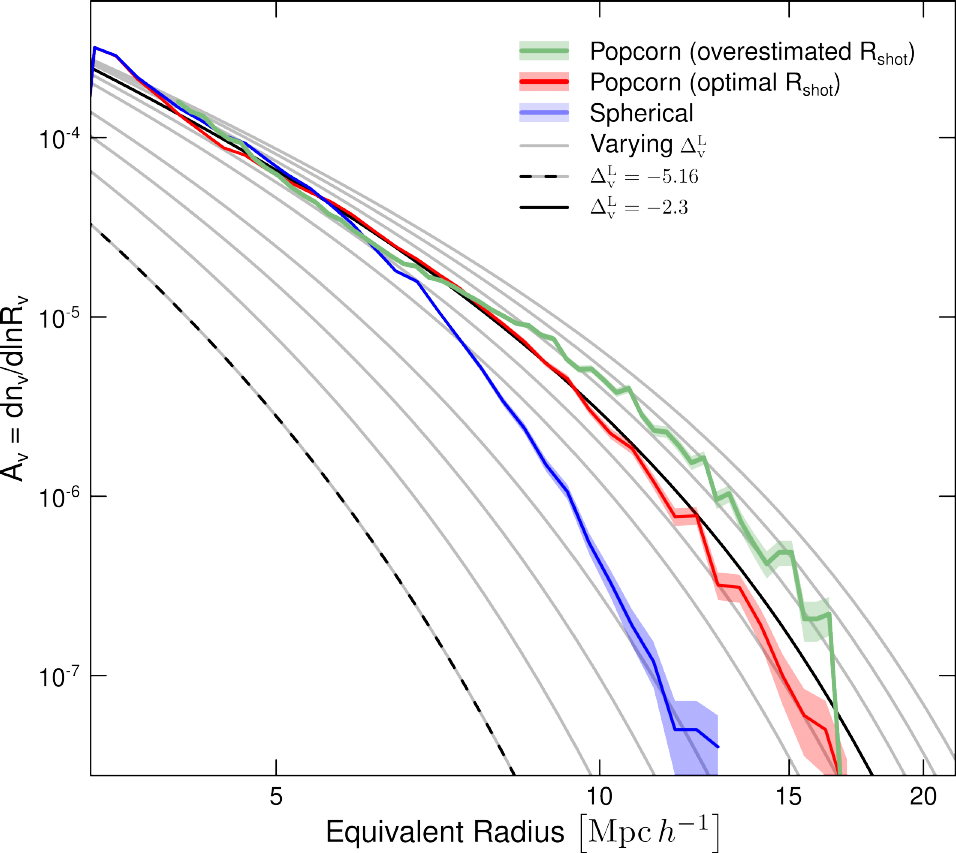}
    \caption{Abundance of popcorn voids identified in the dark matter 
    particle field as a function of their equivalent radius. The results are
    shown for two different values of $R_\mathrm{th}$: in red when this parameter
    is equal to the radius of the shot noise, and in green when a larger value is
    used. The abundances of spherical voids are shown in blue. The shaded areas
    correspond to the uncertainties of the shot noise. Solid gray, solid
    black, and dashed black lines correspond to different Vdn excursion-set models with
    varying linear density barrier values.
    }
    \label{fig:abundances_matter}
\end{figure}

As presented in the previous subsection, Fig.~\ref{fig:abundances_matter} shows
the abundance of popcorn voids identified in dark matter particles in
$\Lambda$CDM simulations (solid red line). 
As we mentioned before, the solid blue line corresponds to spherical voids. As
can be seen, spherical voids at the small size end have higher abundances
than the models, while the opposite is observed for large radii. This  is probably due to
the fragmentation problem of spherical void finders when identifying void
regions. 
The black dashed line in the figure represents the predicted void abundance using the Vdn model
with a linear density barrier of $\Delta_\mathrm{v}^\mathrm{L}=-5.16$, which corresponds to an
integrated nonlinear density contrast of $\Delta_\mathrm{v}=-0.9$ in the spherical expansion model 
(See Sect. \ref{sec:vsf_model}).
As can be seen, this prediction significantly underestimates the abundance of spherical
and popcorn voids with 
this specific equivalent integrated density contrast in the simulation box. The solid
grey lines depict the results of the Vdn model using different values of $\Delta_\mathrm{v}^\mathrm{L}$.
Notably, a value close to $\Delta_\mathrm{v}^\mathrm{L}=-2.3$ provides a good fit to
the popcorn abundances. Note that no feasible value of the linear density barrier can be found to obtain
a reasonable fit for the spherical void sample. On the other hand, when the same experiment is
conducted using the SvdW model, the predicted abundances show a steeper slope and vary much more
rapidly (results not shown for brevity) and those model results cannot be reconciled with the 
abundances of spherical or popcorn voids
for any value of $\Delta_\mathrm{v}^\mathrm{L}$.
Similar results are obtained when identifying spherical and popcorn voids using a nonlinear
density contrast of $\Delta_{v}=-0.8$. For this threshold, the spherical evolution model provides
a linear density contrast barrier of $\Delta_\mathrm{v}^\mathrm{L}=-2.8$, which leads to a significant
underestimation of the abundances once again. However, we find that a better fit to the popcorn abundances can be achieved using a higher value of density barrier (in this case, $\Delta_\mathrm{v}^\mathrm{L}=-1.3$).

Since the abundance models fail to reconcile the observed spherical 
void abundances, we can infer that popcorn voids may be more suitable for
studying abundances.
As we anticipated when analysing
the simply test presented in Fig.~\ref{fig:cartoon}, the identification using
free-form integrated densities allows a more complete coverage of the void
regions. This seems to have two effects on the abundance of voids, on the one
hand it alleviates the overabundance of small voids in fragmented regions,
while also allowing a better characterisation of the scale associated with
large voids. It is important to note that here the size of the popcorn object
is defined by its volume, i.e. its equivalent radius, and this scale is assumed
to be comparable to the radius scale used in excursion-set theory. This
approach is different from that used in \citet{2013Jennings} and
\citet{2016Marulli}, where the voids identified in the density field are
associated with the largest integrated density sphere within the void object.

It is worth to emphasise that popcorn void abundances can be modelled by using higher
values of the linear density barrier, rather than those expected in the spherical expansion model.
These deviations from the spherical expansion model may be due to various factors, including deviations
from the isolation assumption. The non-spherical nature of void regions arises from their interaction
with other structures \citep[see for instance][]{2016Ceccarelli}. Furthermore, any differences
between the nonlinear evolution predicted by
the spherical model and the dynamics of voids observed in the numerical simulation could also
contribute to these deviations in the relation between $\Delta_\mathrm{v}^\mathrm{L}$ and $\Delta_\mathrm{v}$.
Additionally to this last parameter, the VSF models are dependent also on cosmological parameters
and the critical density for collapsing objects (here we adopt $\Delta_c=1.6$).

\begin{figure*}
    \includegraphics[width=2\columnwidth]{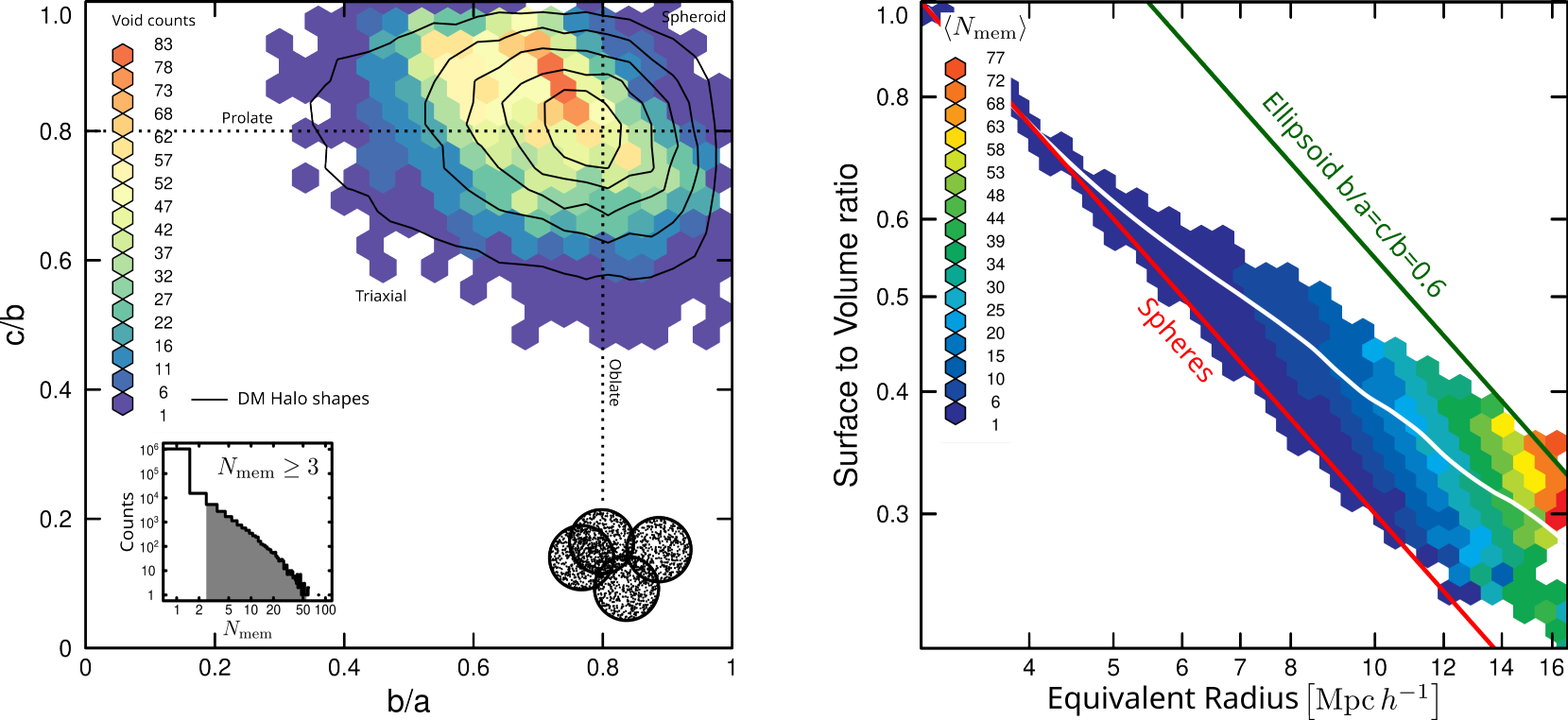}
    \caption{\textit{Left panel:} Distribution of popcorn objects with three or
	more member spheres ($N_\mathrm{mem}\ge 3$) according to the length
	ratio of the semi-axes of the void shape, b/a and c/b. The number of
	objects in each hexagonal bin is indicated by the colours (see the bar
	on the left).  Solid line black contours show the distribution of Dark
	Matter halo shapes \citep[$\mathrm{M}_\mathrm{halo}\ge 10^{12}\msunh$,
	see][]{Paz2006}.  The bottom left box is a histogram of the number of
	member spheres per void, while the grey shaded area corresponds to the
	sample in the left panel of the figure.  \textit{Right panel:} Mean
	number of spherical members of popcorn voids $\langle
	N_\mathrm{mem}\rangle$ in hexagonal bins of equivalent radius and ratio
	of surface area to volume. Both axes are in logarithmic scale, the red
	and green solid lines correspond to the expected surface area to volume
	ratios for spheres and triaxial ellipsoids with b/a = c/b = $0.6$ as a
	function of their size. The solid white line corresponds to the void
	average behaviour. }
    \label{fig:shapes_surf_to_vol}
\end{figure*}

In Fig.~\ref{fig:shapes_surf_to_vol}, we present an analysis of the shape,
volume, and surface of the popcorn objects identified in the matter field of
sim-2. In the case of the shape analysis, we use popcorn voids with three or
more member spheres ($N_\mathrm{mem}\ge 3$). As described above (see
Section~\ref{sec:algorithm}), the \textsc{arvo} library provides the surface
area and volume of the union of spheres using an analytical approach. To
analyse the shape of the popcorn objects, we calculate the shape tensor of the
region defined by the union of the member spheres using a Monte Carlo approach.
Inside each sphere, we randomly allocate $N_\mathrm{ran}$ points, assigning
each one a weight or mass, $m_\mathrm{k}$, given by the inverse of the number
of intersecting spheres at its location, that is
$1/N_\mathrm{sph}({x_1}^\mathrm{k},{ x_2}^\mathrm{k},{x_3}^\mathrm{k})$ (where
the location of the point is expressed through its Cartesian coordinates). An
example of such a random set in a popcorn void is shown in the lower right
schematic drawing in the left panel of Fig.~\ref{fig:shapes_surf_to_vol}. This
set of $N_t=N_\mathrm{ran}N_\mathrm{mem}$ random points weighted in this way is
equivalent to a uniform mass distribution within the volume of the popcorn. We
define then a Monte Carlo estimate of the shape tensor of the region as:
\begin{equation}
    I_\mathrm{ij}:=\frac{1}{N_t}\sum_\mathrm{k=0}^{N_t}m_\mathrm{k}
	\left({x_\mathrm{i}}^\mathrm{k}-\bar{x}_\mathrm{i}\right)
	\left({x_\mathrm{j}}^\mathrm{k}-\bar{x}_\mathrm{j}\right)
\end{equation}
where $\bar{x}_\mathrm{i}$ are the components of the geometric centre of the
region:
\begin{equation}
    \bar{x}_\mathrm{i}:=\frac{1}{N_t}
	\sum_\mathrm{k=0}^{N_t}m_\mathrm{k}{x_\mathrm{i}}^\mathrm{k}\nonumber\,.
\end{equation}
The eigenvalues of this tensor are the square of the lengths of the semiaxes
(a, b, c with, a$\ge$b$\ge$c) of the characteristic ellipsoid that best
approximates the popcorn volume. The b/a and c/b quotients allow us an analysis
of the ellipsoidal figures of best fit 
\citep[see for example][and their references]{Frenk1988,Paz2006,Gonzalez2021,Gu2022}
of the objects and their classification in spheroids (b/a$\simeq$c/b$\simeq
1$), prolate type (i.e.  elongated, 1$\simeq$c/b>b/a), oblate type (i.e
flattened 1$\simeq$b/a>c/b) and triaxial (similar ratio between semiaxes
lengths, b/a$\simeq$c/b). As can be seen from the above reference list of
previous works, this type of shape analysis has been widely applied in the
analysis of dark matter haloes, galaxy groups, and galaxy cluster shapes. 

In the left panel of Fig.~\ref{fig:shapes_surf_to_vol}, we include with solid
black lines the isocontours of the shape distribution for dark matter haloes
with masses greater than $10^{12}\msunh$ as presented in \citet{Paz2006}. In
that work we present an extensive analysis of the effect of shot noise in the
determination of halo and galaxy group shapes using shape tensors. In  our
case, shot noise is not an issue due to the large number of random points used,
however we are limited by the number of member spheres of each popcorn object
($N_\mathrm{mem}$). Popcorn voids with only one sphere are, by definition, in
the upper right corner of this figure, while objects with only two spheres are,
by definition, perfect prolate ellipsoids (collapsed on the axis given by
c/b$=1$). The number of member spheres follows a power law  distribution (see
lower left inset of the left panel of Fig.~\ref{fig:shapes_surf_to_vol}) and
more than $98\%$ of the objects have only one or two member spheres, being
mostly small objects ($95\%$ have sizes below $7\mpch$). Given a total sample
of 1065912 popcorn objects identified in sim-2, 1051176 voids have one or two
member spheres, while 14736 objects have three or more member spheres (grey
shaded area in the figure inset). We restrict the shape analysis to those
popcorn voids that have three or more members. With dotted lines at b/a$=0.8$
and c/b$=0.8$, we have divided the space of the axis ratios into four regions,
as indicated in the figure: spheroidal, prolate-like, oblate-like, and
triaxial.  As can be seen, the peak of the shape distribution of void regions
lays in the prolate region of the diagram. Compared to halo shapes, voids
exhibit a similar behaviour, presenting mostly triaxial shapes with a prolate
trend, more pronounced in void than in dark matter haloes. 

In the right panel of Fig.~\ref{fig:shapes_surf_to_vol}, we show the mean
number of sphere members in hexagonal bins of equivalent radius and ratio of
surface area to volume. The surface area to volume ratio (usually indicated as
sa/vol or SA:V in chemistry and physics) can give us an idea of the compactness
of popcorn shapes. The red line corresponds to the sa/vol ratio of spheres as a
function of their radius, that is, a power law with index -1 (for spheres
sa/vol$=3/R_\mathrm{v}$). In the case of triaxial ellipsoids, similar
behaviours are obtained, almost the same index but with different numerical
factors, that is, a straight line parallel to sa/vol of spheres in the log-log
space of the figure. An extreme case seems to be the triaxial ellipsoid with
b/a=c/b=0.6, whose sa/vol ratio is indicated by a solid green line. The
behaviour of popcorn objects seems to follow an intermediate behaviour between
spheres and triaxial ellipsoids. This fact can be understood as an indication
that the topology of popcorn voids is somewhat close to their ellipsoidal
figures of best fit. A complex behaviour or a different compactness of popcorn
regions compared to ellipsoids should result in quite different sa/vol ratios
as a function of size, and not the intermediate behaviour displayed in the
figure. The trend of the average number of spherical members in the hexagonal
binning scheme of the figure (the visible colour gradient), seems to indicate
that depending on the number of members, the sa/vol ratio behaves like
ellipsoids with increasing axis ratio.

\section{Cosmological test using Void abundances}
\label{sec:abundances}

In this section we present an analysis of the void abundances measured in a
biased tracer field. More specifically, we study the void size function (VSF)
of identified objects in the distribution of dark matter haloes and how this
function depends on cosmology and redshift-space distortions. We also show how
the modelling presented in \citet{Correa2} to account for geometric and
redshift space distortions on abundances of spherical voids can also be applied
in our context of integrated density free-form voids. As in that previous work,
we define geometric distortions (GD) as those that arise when assuming a
fiducial cosmology (but different from the underlying one) when transforming
coordinate measurements (position angles and redshifts) to comoving
coordinates. This is the concept behind any AP test, if an object is assumed to
be the same physical length along the line of sight (LOS) and in the plane of
the sky (POS), any distortion that produces a change in the aspect ratio
between these directions can be used to derive the underlying cosmological
parameters. For further details the interested reader can see Section~2 of
\citet{Correa1}. On the other hand, the peculiar velocities introduce
deviations in the estimation of the distances to galaxies by using the Hubble's
law. These redshfit space distortions (RSDs) in expanding voids result in
apparently larger void regions relative to their size in the underlying
three-dimensional space \citep[see][and references there in]{Correa2}. Finally,
combining the correction factors derived by \citet{Correa2} with excursion-set
theory models of the VSF, we present a cosmological test applicable to the
abundance of voids identified in the biased tracer field in redshift space.

\subsection{Abundance of voids in biased tracer field and Alcock-Paczy\'nski distortions}
\label{sec:abundance_distortions}

\begin{figure*}
    \centering
	\includegraphics[width=2\columnwidth]{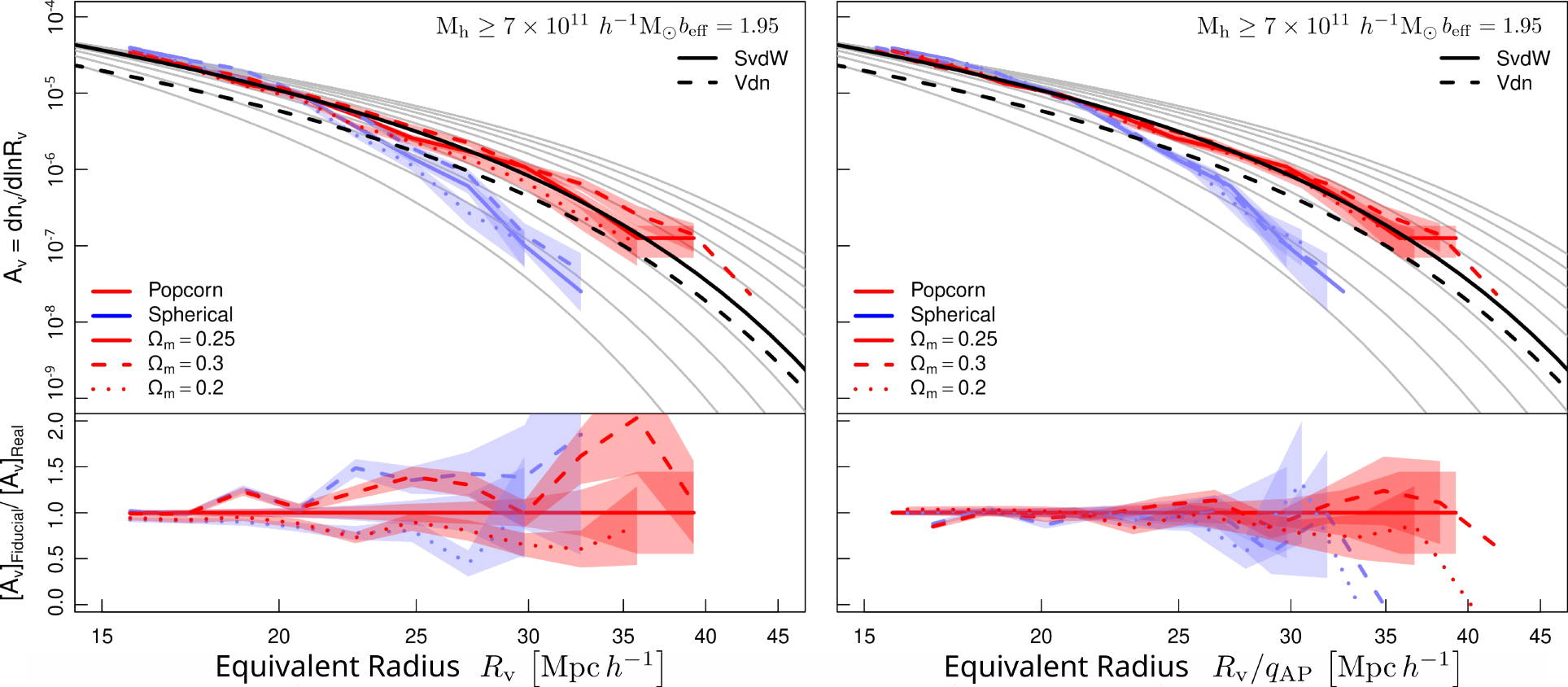}\\
	\vspace{0.3cm}
	\includegraphics[width=2\columnwidth]{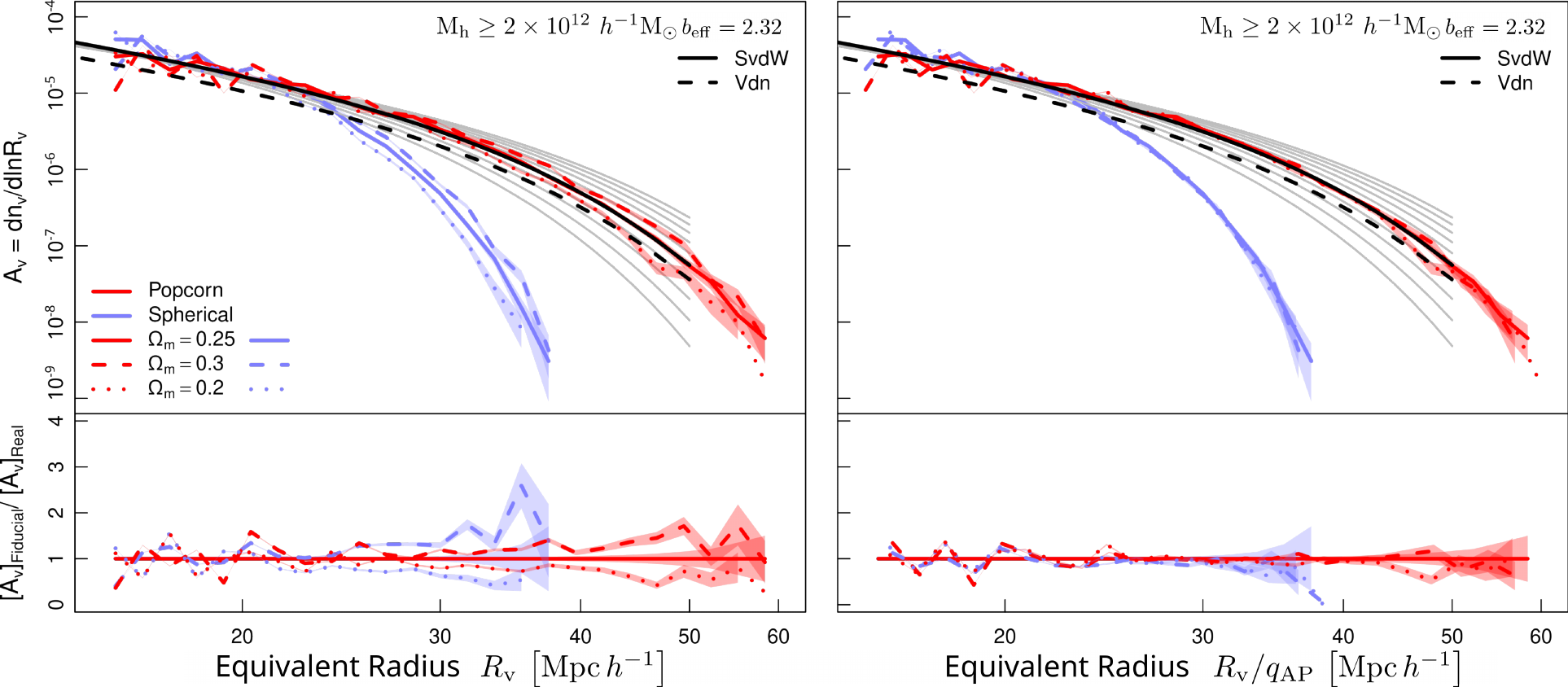}
    \caption{
    \textit{Left panels:} Abundance of spherical voids (solid blue line) and
    popcorn voids (solid red line) identified using the real-space position of
    dark matter haloes with masses greater than $7\times10^{11} \msunh$
    (upper panels) and $2\times10^{12} \msunh$ (lower panels) at 
    $z =0.51$ as a function of the equivalent radius ($\mathrm{A}_\mathrm{v}$).
    The dashed (dotted) line corresponds to the voids identified in the boxes
    with a higher (lower) value of the matter density parameter in the assumed
    fiducial cosmology. In each panel, the quotient between the abundances in
    fiducial and real cosmology is shown below. \textit{Right panels:} The void
    abundances corresponding to the left panels as a function of the equivalent
    void radius corrected for geometric distortions with the factor $\qap$ as
    defined in the text.
    }
    \label{fig:abundancesAP_2cuts}
\end{figure*}

In the left panels in Fig.~\ref{fig:abundancesAP_2cuts}, we show the abundance
of spherical voids (blue solid line) and popcorn voids (red solid line)
identified using the real-space position of dark matter haloes at $z =0.51$.
Haloes were taken as tracers from two simulation boxes, sim-2 and sim-3, with
masses in the range of $M_{\rm halo} \ge 7\times10^{11} \msunh$ and $M_{\rm
halo}\ge 2\times10^{12} \msunh$ (top and bottom panels, as indicated in the
figure key). Similar results are obtained using different halo mass cuts in
both simulation boxes. Using the same colour code (blue for SV and red for
popcorn voids) with dashed and dotted lines, we show the corresponding
abundance of objects identified in boxes with a different fiducial cosmology
than the underlying one. In detail, the dotted and dashed line correspond to
the abundances obtained when the halo coordinates are transformed from true
position to angles and redshifts, assuming a true cosmology, and then
immediately transformed back to comoving space, but this time assuming a
fiducial cosmology with a lower or higher value of $\Omega_\mathrm{m}$. For
this procedure we assume a distant observer and a mean redshift of $z=0.51$.
This transformation of coordinates from real space to observable space and then
back to a comoving space but assuming an incorrect cosmology introduces a
pattern of distortion (GD) in the relative distances along the POS and LOS
directions from the point of view of the observer 
\citep[for a detailed description of geometric distortions see][]{Correa1,Correa2}.
The void size functions on the right panels of
Fig.~\ref{fig:abundancesAP_2cuts} take the same ordinate values as those on the
left, but their arguments (the abscissas $R_\mathrm{v}$) are corrected of GD by
dividing them by the following factor \citep[as derived in][]{Correa2}:
\begin{equation}
   \qap = \sqrt[3]{\left(\frac{D_{\rm M}^{\rm fid}(z)}{D_{\rm M}^{\rm true}(z)}\right)^2 \frac{H_{\rm true}(z)}{H_{\rm fid}(z)}}.
   \label{eq:qap}
\end{equation}
As can be seen, the $\qap$ factor correctly captures the effect of geometric
distortion of void sizes. It depends on the comoving angular diameter distance
at the sample mean redshift ($z$) in the fiducial and underlying cosmologies,
$D_{\rm M}^{\rm fid}$  and $D_{\rm M}^ {\rm true }(z)$, respectively, as well
as in the Hubble parameter for both cosmologies, i.e. $H_{\rm true}$ and
$H_{\rm fid}$. It is not surprising that the same factor can correct the
volumes of spherical and free-form voids, since geometric distortions at a
given fixed mean redshift take the form of a linear coordinate transformation
and $\qap$ is its Jacobian determinant.

In Fig.~\ref{fig:abundancesAP_2cuts} we also include in all panels the
corresponding results of the excursion-set models of \citet{2013Jennings} and
\citet{SvdW} using thick dashed and solid black lines, labeled as Vdn and SvdW
in the figure key, respectively. As was introduced in Sect. \ref{sec:vsf_model},
to predict void abundance, it is necessary to convert the integrated density 
threshold used in void identification (for both
spherical and popcorn voids) from $\Delta_\mathrm{v,g}=-0.9$ over the tracer
sample to a density threshold in the matter field by means of a bias factor, ie
$\Delta=\Delta_\mathrm{v,g}/b_\mathrm{eff}$ 
\citep[as described in][]{2017RonconiMarulli}.  The bias factors
$b_\mathrm{eff}=1.95$ and $b_\mathrm{eff}=2.32$ have been used for the low and
high mass halo samples, respectively (upper and lower panels as indicated in
Fig.~\ref{fig:abundancesAP_2cuts}) to fit the results. 

As can be seen, in this case of biased tracer samples, the model that best fits
the simulation data is the SvdW. By definition, the Vdn model always predicts a
lower amplitude at small scales than SvdW (and consequently lower than
simulation data) due to the volume fraction conservation constraint, regardless
of the tracer bias used.  This differs from the results on abundances in the
matter field presented in Section~\ref{sec:algorithm} and may be related with
an effect of tracer bias on the volume conservation constraint assumed in the
Vdn model. In addition, the reduction in the number of tracers may be affecting
void sizes in a similar manner to the shot noise effect illustrated in Figure
\ref{fig:testshot}, however, this effect may be compounded with the tracer
bias.  In the \citet{2013Jennings} work, the authors also compare the
abundances predicted by the Vdn model with void objects obtained from a
different void finder than the one used here. Despite this, they also find
discrepancies between the model and the data. As these authors explain, this
problem may be related to the fact that the voids identified in the halo
samples are not simply related to the void samples identified in matter, a fact
that is common to all void finders. A one-to-one relationship between halo
voids and matter voids is not feasible, the time evolution of halo voids
depends on the definition of the halo sample (the mass cut for instance). Based
on halo mass and assembly history, a void defined in a given halo tracer field
at a given redshift does not uniquely correlate with a precursor void at an
earlier time.  In this work, we simply apply the abundance models (Vdn and
SvdW) by transforming the nonlinear void density contrast threshold (used in
the popcorn identification) in the halo field to an equivalent value in the
matter field through a free bias parameter, used to fit the data.  It is
important to note that this bias parameter is also encoding by definition
possible deviations in the spherical expansion model, as was described in
section \ref{subsect:prop}.  Our goal is to find a suitable model for the
inference of cosmological parameters from the abundance of popcorn voids.
Although it would be very important to have an adapted model for voids
identified in biased tracer samples, we define this issue as outside the scope
of this work.

Going back to the results in Fig.~\ref{fig:abundancesAP_2cuts}, at small sizes
and regardless of the assumed bias, the SvdW model better predicts void
abundance for both spherical and popcorn objects. The solid grey lines in the
figure correspond to $0.1$ variations in the bias factors, from
$b_\mathrm{eff}-0.5$ to $b_\mathrm{eff}+0.5$. As we mentioned before and can be
seen in the figure, the abundances at small radii are insensitive to the
assumed bias \citep[as it is mentioned in][]{2013Jennings}.  On the other hand,
the behaviour of the VSF models at large sizes seems incompatible with the
abundance of spherical voids. In contrast, the SvdW model fits well with the
abundance of popcorn objects for a proper bias parameter. It is important to
emphasise the fact that here the bias factor is used as a nuisance fitting
parameter and is not necessarily related to the bias factor between the
clustering measures, or between matter and halo velocity fields.  For the halo
samples analyzed in this section, namely those with $M_{\rm halo} \ge
7\times10^{11} \msunh$ and $M_{\rm halo} \ge 2\times10^{12} \msunh$, the
expected values of the clustering bias at $z=0.51$ are $0.91$ and $1.05$,
respectively, as computed using the formulae presented in \citet{2010Tinker}.
As can be seen, these values for clustering bias are much smaller than the
corresponding best-fit $b_{\rm eff}$ values. We see larger differences between
the effective bias required to fit popcorn abundances and the clustering bias
compared to similar studies using other void finders
\citep{abundance_bias_contarini, bias_pollina1, bias_pollina2}.  The question
of tracer bias around void regions remains open and is an active research
topic.  For further information, we recommend also reading the following works:
\cite{bias_chan2, density_bias_fang, bias_chan1, bias_schuster, bias_chan3}.

\subsection{Model fitting of void abundances in redshfit space}
\label{sec:test}

\begin{figure*}
    \includegraphics[width=2\columnwidth]{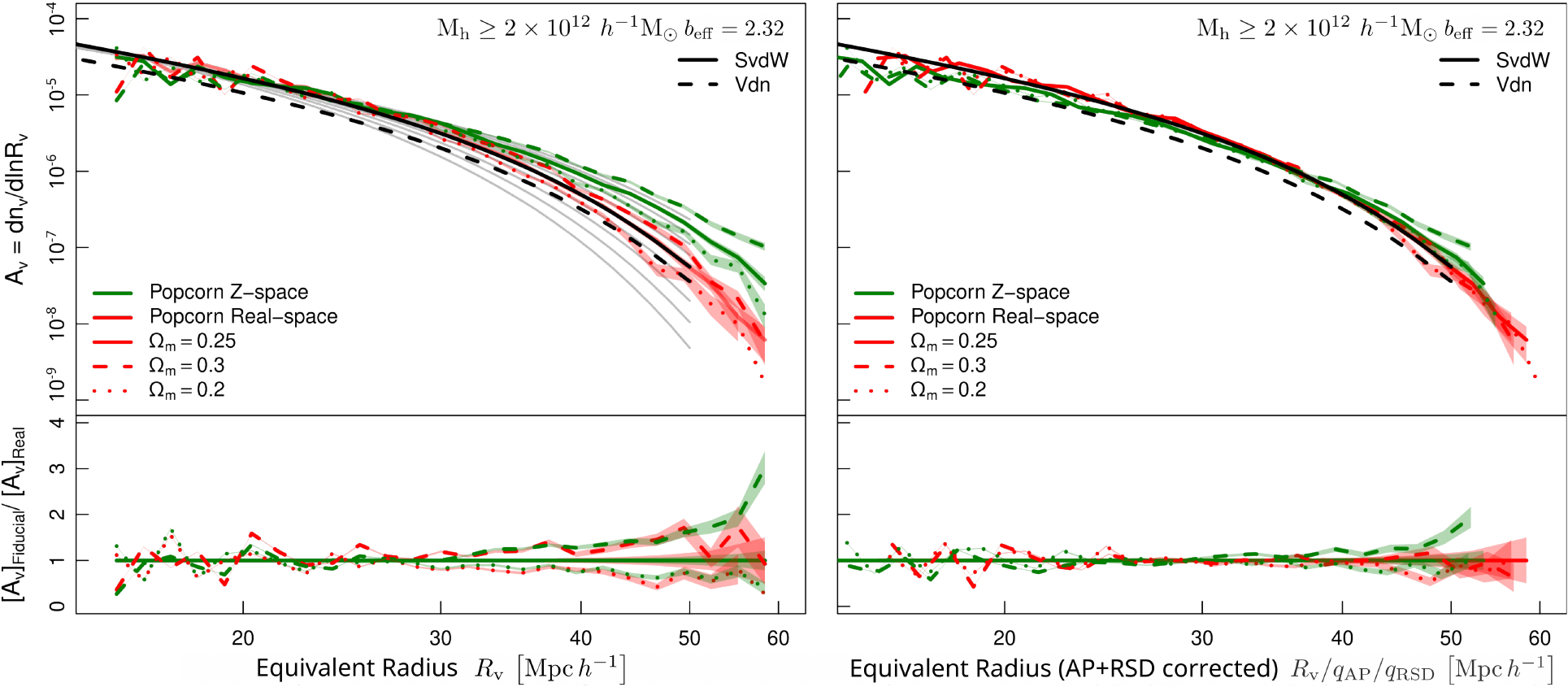}
    \caption{ \textit{Left panel:} Abundance of popcorn voids in real and
	redshift space, solid red and green lines, respectively, identified on
	dark matter haloes with masses greater than $2\times10^{12} \msunh$ at
	$z =0.51$ as a function of the equivalent radius
	($\mathrm{A}_\mathrm{v}$). The dashed (dotted) line corresponds to the
	voids identified in the boxes with a higher (lower) value of the
	density parameter in the assumed fiducial cosmology.  The quotient
	between the abundances in fiducial and real cosmology is shown below.
	\textit{Right panel:} The void abundances corresponding to the left
	panels as a function of the equivalent void radius corrected for
	geometric and redshift-space distortions with the $\qap$ and $\qrsd$
	factors as defined in the text.}
    \label{fig:abundancesZ}
\end{figure*}

So far, we have addressed the effect of geometric distortions (the
Alcock-Paczy\'nski effect) and tracer bias on void abundance, now we will check
if the treatment of redshift-space distortions presented in \citet{Correa2} can
be applied in the context of popcorn voids. In an analysis similar to that
presented in our previous works, using the distant observer approach, we add
the peculiar velocity component in the LOS direction as an additional Doppler
term to the cosmic redshift. This introduces a redshift-space distortion (RSD)
pattern at the halo positions \citep[see for example][]{clues2}. 
Being voids regions of negative density contrast
a positive divergence of the velocity
field is expected, and thus the apparent position of the tracer haloes in
redshift space is shifted toward the outskirts of the void. Therefore, the size
of voids identified in redshift surveys tend to be larger than in real space,
resulting in a right-shifted VSF along the $R_\mathrm{v}$ axis.

In \citet{Correa2} we have derived a correction factor for the RSD in the
radius of spherical voids from linear theory:
\begin{equation}
    \qrsd = 1 - \frac{1}{3} \delta R_{\rm v} \beta(z) \Delta_\mathrm{v,g}.
    \label{eq:qrsd}
\end{equation}
where ${\beta}$ is the growth rate of structures over a bias factor 
\citep[see for instance][and references there in]{clues2,Hamaus_2015} and
$\delta R_{\rm v}$ is a parameter to quantify the variation on the predicted
expansion of the void radius. This factor is derived from the expected
expansion of a hypothetical underdense spherical region, of radius
$R_\mathrm{v}$, due to RSD.  In that context, the elongation produced by RSD
along LOS distorts this sphere into an ellipsoid in redshift space, with a
major semi-axis greater than $R_\mathrm{v}$ and intermediate and minor
semi-axes equal to $R_\mathrm{v}$.  When a spherical void finder is used, this
region is associated with a spherical void object with a radius between the
semi-major axis and the radius in real space. Therefore, the derivation of a
theoretical value of $\delta R_{\rm v}$ will depend on the details of the
spherical void profile around the scale associated with the integrated density
threshold ($\Delta_\mathrm{v,g}$).  In \citet{Correa2} we have derived for
spherical voids a mean value of $\delta R_{\rm v}$ between $0.5$ and $0.3$,
depending on the size of the void.  However, it is important to stress the fact
that in the case of popcorn voids we have chosen to study its abundances as a
function of its equivalent radius, that is the radius of the sphere with a
volume equivalent to the one of the identified object. In this work we will
correct the equivalent popcorn void radius by dividing it with the $\qrsd$
factor defined in Eq.~(\ref{eq:qrsd}).  However, the interpretation of the role
of $\delta R_{\rm v}$ has to be different. The popcorn void finder is designed
to define free-form objects, so a void region can be considered to a first
approximation as an ellipsoidal region, with an arbitrary orientation, assumed
to be uniform due to cosmic isotropy. It can be assumed that the shape
parameters of these regions follow the distributions shown in
Fig.~\ref{fig:shapes_surf_to_vol}. In the context of popcorn voids, $\delta
R_\mathrm{v}$ depends on the orientation of the void, the details of the
velocity field around the void region (not necessarily well described by the
model of spherical linear expansion) and the average of the different shapes in
a bin of equivalent radius. In this work, $\delta R_{\rm v}$ will be used as a
free parameter to adjust the RSD of the popcorn void sizes.

Since $\beta$ is a parameter defined in the spherical expansion model, we
derive it from spherical voids, following the same procedure as in
\citet{Correa1} and \citet{Correa2}. In a tracer sample of haloes with masses
greater than $2\times10^{12} \msunh$ at $z =0.51$, we have measured the radial
velocity and density profiles of spherical voids, finding a $\beta $ value of
$0.54$ in a range of sizes from $15$ to $40\mpch$.  As we will see below,
popcorn voids are well characterised with an RSD parameter of $\qrsd=1.12$,
which corresponds to an increase in radius of $\delta R_\mathrm{v}=0.74$.  

In Fig.~\ref{fig:abundancesZ} we show the abundance of popcorn voids in real
space (solid red lines) and redshift space (solid green line) identified using
dark matter haloes with masses greater than $2 \times10^{12} \msunh$ at $z
=0.51$. In the left panel, the abundances are shown as a function of the
equivalent radius of the void (as defined by Eq.~\ref{eq:eqvrad}) calculated
from the volume of popcorn objects identified in real or redshift space,
depending on the case considered. As in the previous analysis, dashed and
dotted lines correspond to voids identified after the coordinates of each halo
are affected by geometric and redshfit distortions. This is done by
transforming halo coordinates from real to observable space, using the
simulation's underlying cosmology, adding a component of Doppler shift along
the LOS, and finally transformed back to comoving space but using a fiducial
cosmology with a different value of the density parameter. The quotient between
abundances in fiducial and real cosmology is shown in the bottom plot of each
panel. In the right panel we show the abundance of popcorn objects once their
volumes are divided by the correction factors $\qap$ and $\qrsd$. The shaded
areas correspond to Possion noise standard deviations. As can be seen the
application of both factors successfully correct the void size functions from
geometric and redshift-space distortions. The abundances shown in the right
panel of Fig.~\ref{fig:abundancesZ} are indistinguishable in almost all scales,
showing marginally differences between them at large sizes ($R_\mathrm{v}>
40\mpch$), where the cosmic variance is more important. As in the
Fig.~\ref{fig:abundancesAP_2cuts}, the SvdW model with a bias factor of $2.32$
seems to be good fit. 

\begin{figure*}
    \includegraphics[width=2\columnwidth]{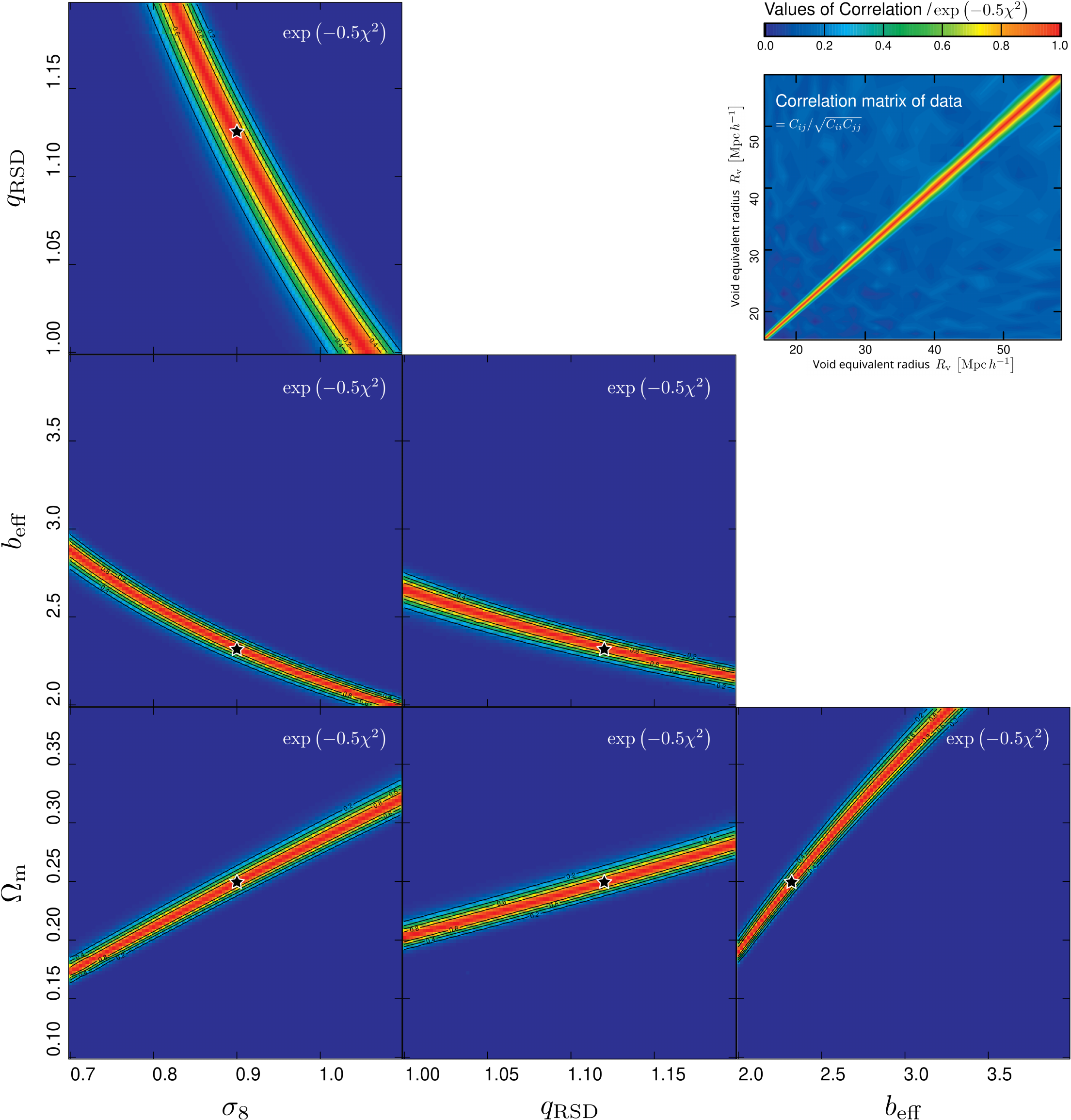}
    \caption{
    Different two-dimensional slices of the likelihood function of
    the cosmological test on the abundance of popcorn voids (top left
    and both bottom panels). The parameter space is composed of the
    matter density parameter ($\Omega_\mathrm{m}$), the RMS linear
    perturbation variance at $8\mpch$ ($\sigma_8$), the effective bias
    of the tracers ($b_\mathrm{eff}$, used in the excursion-set model)
    and the redshift-space expansion factor of void sizes, $\qrsd$.
    Void abundances were measured assuming a fiducial cosmology with
    a lower value of $\Omega_\mathrm{m}=0.2$ and a halo sample with
    $M_\mathrm{halo}\ge 2\times10^{12} \msunh$ in the
    redshift space at $z =0.51$ (see text). The top right panel shows
    the correlation matrix of void abundance measurements at
    different equivalent void radii. Both quantities, correlation and
    likelihood vary between 0 and 1, and are indicated following the
    colour table from blue to red as shown in the colour bar in the figure. 
    The black star indicates the expected target parameter set
    for this test.
    }
    \label{fig:likelihood}
\end{figure*}

So far we have described how to correct void abundance measurements for
redshift space distortions and how to model the Alcock-Paczy\'nski effect. We
have also shown a good apparent concordance of these measurements with
excursion-set models, after applying the appropriate corrections. Now we will
analyse the model fit statistics, in particular the combination of excursion-set
models with the factors $\qrsd$ and $\qap$ to obtain information on
cosmological parameters.  We will limit ourselves to the matter density
parameter and the root mean square variance of linear perturbations to show the
potential of the abundance of voids as cosmological test, leaving the analysis
of a larger parameter space for future work. In Fig.~\ref{fig:likelihood}, we
show the likelihood function for the evaluation of the SvdW model for the
abundance of popcorn voids. As above, the voids are identified using haloes in
redshift space and assuming a fiducial cosmology, in this case with a lower
value of the matter density parameter, i.e. the dotted curves shown in
Fig.~\ref{fig:abundancesZ}. Identical results are obtained using higher values
of the density parameter, indicating that the test appears to be independent of
the assumed fiducial cosmology. 

The parameter space explored consists of the matter density parameter
$\Omega_\mathrm{m}$, the root mean square variance of linear fluctuations at
$8\mpch$ $\sigma_8$, an effective bias $b_\mathrm{eff}$, and the correction
factor of redshift-space distortions $\qrsd$. The effective bias is used to
convert the halo density contrast threshold used in popcorn identification,
$\Delta_\mathrm{v}=-0.9$, to an equivalent nonlinear density contrast in the
matter field, i.e.  $\Delta_\mathrm{NL}=\Delta_\mathrm{v}/b_\mathrm{eff}$. This
value is then transformed into a linear value by means of the fitting formula
presented in \citet[][, see Sect. \ref{sec:vsf_model}]{1994Bernardeau}.  In
this way, the linear density contrast in the matter field, $\Delta_\mathrm{L}$,
corresponding to the density threshold used in void identification, is used as
the void density barrier in the excursion-set models.  See
Sec.~\ref{sec:vsf_model} for more details.  The matter density parameter plays
a role in the evaluation of excursion-set models (through the matter spectrum
in the barrier statistics) and in the factor $\qap$. The value $\sigma_8$
determines the amplitude of the power spectrum that then directly impacts the
VSF. Finally, we have chosen to fit the factor $\qrsd$ instead of $\delta
R_\mathrm{v}$ to decouple the bias factor in the velocity field (which appears
in the definition of $\beta$, Eq.~\ref{eq:qrsd}) from the effective bias
($b_\mathrm{eff}$) used in the computation of the density barrier in the SvdW
model. The measurement of the bias in the velocity field by means of the
velocity curves of spherical voids as presented before, gives us inconsistent
values of beta with $b_\mathrm{eff}$. This could be an indication of the
inadequacy of the linear spherical expansion model for the velocity field
around popcorn voids or the need for an improvement in the treatment of biased
tracer samples in excursion-set theory.  Moreover, it is possible that the
relationship between the linear barrier and the non-linear integrated density
of popcorn objects may not be straightforward.  Furthermore, the spherical
model may not provide a good model for non-linear evolution of non-spherical
void regions.  Here we treat the $\qrsd$ factor as a nuisance parameter,
leaving a detailed study of the velocity and density profiles of popcorn voids
for future work.  To extract the most information from the cosmological test
presented here, an adequate model (beyond the scope of this work) is required
that describes the relationship between the different density contrasts
involved. These include the linear barrier in excursion-set models, the
non-linear value used to identify voids in biased tracer samples, and the
density contrast of matter, which is related to the velocity expansion of the
region and its distortions in redshift space.  An elliptical or non-isolated
non-linear model for void evolution is a crucial ingredient to allow void tests
to extract the maximum cosmological information.

Lastly, we will analyse the likelihood of the fits presented above in the
parameter space of $\Omega_m$, $\sigma_8$, $\qrsd$ and $b_\mathrm{eff}$.  As a
first step, we need to estimate the covariance of the void abundance
measurements. This is done using the jackknife resampling technique. We divide
the void sample into $N_\mathrm{jack}=1000$ subsets of
$N_\mathrm{v}/N_\mathrm{jack}$ non-repeating objects. Each jackknife measure of
the void abundance, denoted by ${A_\mathrm{v}}^k$ where
$k=1,\dots,N_\mathrm{jack}$, is made over the entire sample but removing one of
these random subsets of the calculation, i.e. using
$N_\mathrm{v}(N_\mathrm{jack}-1)/N_\mathrm{jack}$ objects.  Given the set of
jackknife realisations, we are allowed to define the mean jackknife measure in
the $i$-th bin in void sizes as:
\begin{equation}
    \overline{A_\mathrm{v}}_i=\frac{1}{N_\mathrm{jack}}
	\sum_{k}^{N_\mathrm{jack}}{A_\mathrm{v}}^k_i\nonumber
\end{equation}
and consequently estimate the covariance of the void abundances as follows:
\begin{equation}
    C_{ij}=\frac{N_\mathrm{jack}-1}{N_\mathrm{jack}}\sum_{k}^{N_\mathrm{jack}}
    \left({A_\mathrm{v}}^k_i-\overline{A_\mathrm{v}}_i\right)
    \left({A_\mathrm{v}}^k_j-\overline{A_\mathrm{v}}_j\right)
    \nonumber\,.
\end{equation}
Given this data covariance and an array of the differences between the
abundance measure (using the whole void sample) and the model,
$\delta\mathbf{A}=\mathbf{A}_\mathrm{model}\left(\Omega_m,\qrsd,b_\mathrm{eff},\sigma_8\right)-\mathbf{A}_\mathrm{v}$,
the likelihood function is defined as usual:
\begin{align}
    \chi^2&=\sum_{ij}{C^{-1}}_{ij}\delta{A_i}\delta{A_j}\nonumber\\
    \mathcal{L}&\propto \left|\mathbf{C^{-1}}\right|^{1/2}\mathrm{exp}
    \left[-\frac{1}{2}\chi^2\right]\,.
    \label{likl}
\end{align}

In Fig.~\ref{fig:likelihood} we show in the top right panel the covariance
matrix, conveniently normalised by the variances in the pair of sizes
associated with the $i,j$ position, in other words the correlation matrix
defined as $C_{ij}/\sqrt{C_{ii}C_{jj}}$.  This normalisation allows us to see
more clearly the structure of the covariance in the measurement of void
abundances. Correlation matrix values range from 0 to 1 and are indicated by a
colour table ranging from blue to red, as indicated in the figure.  The
estimation of $A_\mathrm{v}$ at a given void size is not independent of the
values at near sizes, due to $C_{ij}$ is not diagonal, however as can be seen
is close to it. This allows us the use of different methods to control the
propagation of errors in the likelihood estimation, in particular in the
precision matrix defined as $\mathbf{C^{-1}}$, using for example the covariance
tapering \citep[][]{Paztaper}.  However, we have verified here that, due to the
large box and the number of jackknife resamples used in this work, this type of
cleaning has little effect on the parameter constraints. Nonetheless, in the
case of measurements on observations, this quasi-diagonal structure of the data
covariance could be exploited to control systematic errors or alleviate the
required number of mock catalogues in the analysis.

In Fig.~\ref{fig:likelihood} we also show different two-dimensional slices of
the likelihood function in the parameter space.  As can be seen in all these
projections, the likelihood function shows a high degree of degeneracy for each
pair of parameters. Because of this behaviour, we have chosen to evaluate the
likelihood function on a fine cubic grid in the parameter space instead of
using Monte Carlo likelihood estimation methods such as Markov chains.  The
point marked with a black star indicates the expected parameters: the
simulation box underlying values of $\Omega_\mathrm{m}$ and $\sigma_8$, the
effective bias determined in real-space measurements of void abundances
$b_\mathrm{eff}$, and the $\qrsd$ factor, all obtained using the true cosmology
of the simulation box.  The introduction of geometric and dynamic distortions
in this mock sample of voids does not appear to introduce any bias with respect
to these target values.  In particular, the nuisance parameter $\qrsd$ obtained
in this fit is insensitive to geometric distortions. Indeed, we have repeated
the analysis presented in Fig.~\ref{fig:likelihood}, obtaining the same
best-fit parameters and constraints regardless of the assumed fiducial
cosmology. In this case we show the results for a test run assuming a low value
of $\Omega_\mathrm{m}=0.2$.  Despite the long degeneracy shown in the parameter
space for this cosmological test, the lack of bias with respect to the target
parameters suggests that the popcorn void abundances can be combined with other
probes (as supernovae, BAO, CMB, etc.), to find better cosmological parameter
constraints.  The constraints obtained on the plane defined by
$\Omega_\mathrm{m}$ and $\sigma_8$ are in qualitative agreement with recent
results in the literature for both observations and forecasts
\citep{abundance_bias_contarini,ContariniArxiv1,ContariniArxiv2,Euclid_forecast,2022Pelliciari}.
Notably, the slope of this degeneracy appears to be orthogonal to those derived
from other standard probes. This is an important result, as it helps to break
the degeneracy between these parameters when using joint probes.

\section{Summary and Conclusions}
\label{sec:discus}

In the first part of this work, Section~\ref{sec:algorithm}, we have presented
a new definition of cosmic voids and the accompanying software package where it
is implemented. The code is publicly available and ready to be used in the
context of cosmological simulations, we leave the implementation for use in
galaxy redshift surveys for future work. The popcorn void finder defines voids
as regions of maximum free-form volume with a given integrated density
contrast. Due to these two features, this definition of a void can be naturally
related to predictions of void abundance based on excursion-set theory. The
maximum size of a void defines a unique scale associated with the adopted
density contrast threshold, $\Delta_\mathrm{v}$.  Once a prescription has been
applied to associate nonlinear densities with linear values \citep[see, for
instance, the approach presented by][]{1994Bernardeau}, the quantity
$\Delta_\mathrm{v}$ can be mapped onto a linear barrier that can be used to
compute the crossing statistics on a linear Gaussian density field within the
excursion-set theory.  Although spherical void finders also provide a
characteristic scale for a given $\Delta_\mathrm{v}$, the assumption of a
spherical shape results in what we call a fragmentation problem in void
identification. A single void region, depending on its size and shape, is often
identified as multiple objects.  This results in an artificial increase in the
abundance of small voids, while the opposite occurs for larger regions. The
popcorn void finder algorithm seems to solve this problem. The finder only
depends on two parameters, the integrated density contrast threshold
$\Delta_\mathrm{v}$, and the minimum radius of a popcorn sphere member,
$R_\mathrm{shot}$, however, this last parameter can be fixed. We have shown
that this radius is closely related to the shot noise scale, which is the size
at which an underdense sphere is statistically significant in a given tracer
sample. 

After describing the void finder, we report the first significant finding
of this work: the abundance of popcorn voids identified in the matter density
field can be accurately described using the model proposed by \citet{2013Jennings},
once a suitable linear density threshold is chosen. However, the \citet{SvdW} model
does not provide a viable fit. In contrast, neither of these models provides
a good fit for spherical voids. The agreement between the results obtained from
the popcorn void finder in simulations and the theoretical models critically
depends on the choice of a linear barrier that is generally larger than expected
in the spherical expansion model. This suggests the need to re-evaluate the
relation between linear and nonlinear density contrast in the context of a new,
perhaps non-spherical or non-isolated model.
On the other hand, the accurate estimation of the shot noise radius is crucial 
for obtaining reliable results with the popcorn void finder. The abundance of
voids in the matter field could have practical relevance if extended to the analysis
of void abundances in lensing fields \citep[see, for instance,][]{2021Davies}

We have also analysed the shape of the popcorn voids in the matter field. We
have found that they can be well described as triaxial ellipsoids, their
surface area to volume ratio suggesting that their topology is not much more
complex than that of ellipsoids.  On the other hand, the distribution of
popcorn voids in the axis-ratio space of ellipsoidal shapes is qualitatively
similar to that of dark matter haloes. Voids as rare underdense regions that
have triaxial shapes with predominance of prolate configurations, similar to
what is expected from rare peaks in the matter field.  Furthermore, this
appears to be in qualitative agreement with the peak patch picture developed by
\citet{1996BondMyers}. These authors show that in the context of the
constrained statistics of extremes in a Gaussian field, the matter around the
local extremes tends to be distributed in prolate configurations. In our
opinion, these results originally interpreted in the context of dark matter
haloes can also be applied to void shapes. In a future work we will delve into
the analysis of void shapes and their possible use as cosmological probes The
ellipticity of voids, in real space, have been proposed as probes of dark
energy
\citep[][]{2007ParkLee,2009LeePark,2010Biswas,2011Lavallaz,2012Bos,2015Pisani,2020Rezaei}
as it can affect the shape of voids and therefore their ellipticity
distribution.  On the other hand, in \cite{Correa3}, we showed how the
intrinsic ellipsoidal nature of voids can be detected from RSD analyses of the
VGCF, and the impact of this effect on cosmological tests.

In the second part of this work, Section~\ref{sec:abundances}, we have
presented an analysis of the abundance of popcorn voids in biased tracer
samples, i.e. using different dark matter mass cutoffs, in real and redshift
space. We have shown that the void abundances in dark matter halo samples are
susceptible to geometric distortions due to the assumption of a given fiducial
cosmology. We have also shown that the framework developed by \citet{Correa2}
can be applied in the context of popcorn voids.  Using the factor $\qap$
derived in that work, we have established that the abundance of popcorn voids
can be used to perform an Alcock-Paczy\'nski test.

On the other hand, we also analysed the abundance of identified popcorn voids
in the redshift space. Using the \citet{Correa2} formalism, we have shown that
popcorn void sizes behave qualitatively as expected, void regions are
apparently larger in redshift space. We have also shown that the abundance in
redshift space can also be corrected using a $\qrsd$ factor, as derived in
\citet{Correa2}. However, in the case of popcorn voids, the distortion of sizes
in redshift space tends to be higher compared to what was found for spherical
voids in that earlier work. This could be an indication of differences between
the linear spherical expansion model and the velocity field in neighbouring
regions of the popcorn voids. In a future article we will present a detailed
analysis of the density and velocity field in the popcorn surroundings.

Finally, using the abundance of popcorn voids in the biased tracer field in
redshift space, we have developed a cosmological proof using excursion-set
theory and the \citet{Correa2} framework. As an example of the possible
exploitation of void abundances in cosmological tests, we have analysed the fit
of the model in the parameter space formed by the matter density parameter
$\Omega_\mathrm{m}$, the RMS variance of linear fluctuations $\sigma_8$, both
as parameters of interest, and two nuisance parameters as described below. The
first nuisance parameter is an effective bias, $b_\mathrm{eff}$, used to relate
the halo density contrast threshold, used in the popcorn void finder, to the
nonlinear density contrast in the matter field, used in the excursion-set
model. The second parameter is the $\qrsd$ factor, which is used to correct
sizes due to redshift-space distortions. This last quantity is used as a
nuisance parameter, due to our ignorance of an accurate modelling of the
velocity field around popcorn voids. However, we hope, as in the case of the
linear theory, that this parameter can be related to the growth rate of the
structures and, after correct modelling of the bias in the velocity field, it
can be used to test different gravity models. 

We have found that, in the context of biased tracer samples, the \citet{SvdW}
model appears to be more suitable for describing the abundance of popcorn
voids. The discrepancies with the \citet{2013Jennings} model may be related to
a possible improper treatment of the halo bias in the volume conservation
constraint. Because the aim of this work is not to develop excursion-set
models, but rather to show the potential exploitation of popcorn voids in
cosmological probes, we apply halo bias prescriptions at the most naive level.
As we mentioned before, the halo density contrast threshold is converted to a
nonlinear matter value through the effective bias, then this value is used to
derive a linear density barrier using the prescription of
\citet{1994Bernardeau}.

The likelihood function on the parameter space for the proposed cosmological
test on the abundance of popcorn voids shows large degeneracies. However, the
test shows statistical robustness, that is, there are no biases in the target
parameters: the expected parameters coincide with the best fit parameters. This
situation is common in various cosmological probes, however, joint analysis has
been the key in the era of precision cosmology, reducing parameter constraints
and allowing different paradigms to be discarded.  In a recent paper
\citet{2022Pelliciari} present a cosmological test using the VSF in combination
with the halo mass function. Using a similar methodology,
\citet{Euclid_forecast} provide forecasts for the Euclid mission of the
constraining power of cosmological probes based on the void abundances.  In
those works the cleaning and rescaling method of \citet{2017RonconiMarulli} is
applied over the a void sample identified using \textsc{vide}
\citep[][]{VIDEpaper}.  Even thought the differences in methodology and in the
nuisance parameters, their results suggest the great potential of cosmological
test of void abundances in combination with other cosmological probes.

In our opinion, the proposed cosmological test for galaxy redshift surveys
using popcorn voids, given the tight contours shown in parameter space, the
unbiased recovering of the parameters, and the few nuisance parameters
required, has great potential to improve cosmological parameter constraints.

\section*{Acknowledgements}

This work was partially supported by the Consejo de Investigaciones Científicas
y Técnicas de la República Argentina (CONICET) and the Secretaría de Ciencia y
Técnica de la Universidad Nacional de Córdoba (SeCyT). We thank Dr. Raúl Angulo
for kindly providing us with the halo data from the MXXL simulation.  DJP
specialy thanks to Dr. Jan Busa Sr., Dr. Jan Busa Jr., Dr. Ming-Chya Wu and Dr.
Chin-Kun Hu for the development of the \textsc{arvo} software and their
invaluable help in its use and adaptation.  Numerical calculations were
performed at the computer clusters from the Centro de Cómputo de Alto Desempeño
de la Universidad Nacional de Córdoba (CCAD, \url{http://ccad.unc.edu.ar}).
Plots were made with the \textsc{r} software \citep{Rsoft} and postprocessed
with Inkscape (\url{https://inkscape.org}). We thank the anonymous referee for
carefully reading this manuscript and providing useful comments and suggestions 
that significantly improved this paper.

\section*{Data Availability}

The \textsc{popcorn} and spherical void finders are publicly available under a
MIT licence in the GitLab repository at \giturl. The data underlying this
article will be shared on reasonable request to the corresponding author.



\appendix

\section{The spherical void finder} 
\label{svf_algorithm}

The SVF defines a void object as the non-overlapping sphere of maximum radius
with an integrated density contrast below a desired threshold
$\Delta_\mathrm{v}$ at a given position. Then the first step is to find the
locations to place the candidate spheres, called seeds. To achieve this, we
first grid the simulation volume into regular cubic bins (voxcels) of side
length $\delta l$. To speed up the calculation of the integrated densities, the
total number of particles within each voxcel is computed. We then select
voxcels for seed placement using the following procedure. Centred on each cell,
we count the total number of tracers, $N_t$, using only the voxcels completely
contained within the intersecting volume of spheres of radius, $r_i$, centred
on the edges of the central cell. The radii $r_i$ are taken in increasing
sequence starting at $(1+ 1/\sqrt{2})\delta l$ and ending at
$R_\mathrm{max}=(N_\mathrm{s}+1/\sqrt{2})\delta l$, in $N_\mathrm{s}$ steps of
$\delta l$. Then, for each radius in this sequence, we calculate a lower
estimate, $\Delta_i$, of the density contrast for spheres centred at any
position within the cell. These estimates are given by
$\Delta_i=N_t/(\tfrac{4}{3}\pi {r_i}^3)$. For each voxcel we keep the largest
radius $r_\mathrm{imax}$ that has an estimated density below
$\Delta_\mathrm{v}$. If $r_\mathrm{imax}$ is greater than a given radius of
interest, the cell is seeded following a uniform random distribution. The
number of random points seeded depends on the desired spatial resolution for
the void centres, given the cell size. For each random seed, we compute the
exact density within a sphere of radius $r_\mathrm{imax}$ centred at the
position of the seed. This is done by opening the voxcels partially inside the
sphere and calculating the distances of its particles to the centre. Then the
particles with distances within the radius are added to the total number of
particles in the voxcels contained within the sphere. If the exact density
contrast is greater than $\Delta_\mathrm{v}$, $r_\mathrm{imax}$ is redefined as
the next smallest radius in the radius array ($\mathrm{imax} \rightarrow
\mathrm{imax}-1$) and the density calculation is repeated. This procedure is
repeated until the exact density on the scale $r_\mathrm{imax}$ is below
$\Delta_\mathrm{v}$ or the seed is discarded, that is, the density condition
cannot be met at any radius of the array $r_i$.

At the end of the seeding procedure, each seed is expected to cross the exact
density threshold at a radius intermediate between $r_\mathrm{imax}$ and
$r_{\mathrm{imax}+1}$.  Next, all the particles within the spherical shell
contained between these two radii are ordered according to their distance from
the centre. For each particle, a density is calculated using its distance as
the radius and the total number of contained particles, including itself. Then
the largest sphere with an integrated density contrast below
$\Delta_\mathrm{v}$ is associated with the seed. All overlapping spheres are
then removed, following an order of increasing size, keeping the largest in
each pair of overlapping spheres. At the end of the process we have a set of
non-overlapping spheres covering the low-density regions of the simulation box.
This is the final catalogue of void objects produced by the SVF.

\label{lastpage}
\end{document}